\documentclass[a4paper,12pt]{article}

\usepackage[numbers]{natbib}

\usepackage{amsmath}
\usepackage{amstext}
\usepackage{amsfonts}
\usepackage{amssymb}
\usepackage{slashed}
\usepackage{amscd}
\usepackage{latexsym}
\usepackage{dsfont}
\usepackage{graphicx}
\usepackage{epstopdf}
\usepackage{color}
\usepackage{subfigure}
\usepackage{sidecap}
\usepackage[DIV12]{typearea}
\usepackage[english]{babel}

\newcommand{\nn}{\nonumber}
\newcommand{\Trace}{\text{Tr}}

\newcommand{\Act}{\mathcal{S}}

\newcommand{\psib}{\ensuremath{\bar{\psi}}}
\newcommand{\Psib}{\ensuremath{\bar{\Psi}}}

\newcommand{\hmu}{\hat{\mu}}
\newcommand{\hnu}{\hat{\nu}}

\newcommand{\nablamu}{\nabla_{\mu}}

\newcommand{\tauone}{\tau^{1}}
\newcommand{\tautwo}{\tau^{2}}
\newcommand{\tauthree}{\tau^{3}}

\newcommand{\gammamu}{\gamma_{\mu}}

\newcommand{\gammafive}{\gamma_{5}}

\newcommand{\Umux}{U_{\mu}(x)}
\newcommand{\Umuxmad}{U_{\mu}(x-a\hmu)^{\dagger}}

\newcommand{\mnull}{\ensuremath{m_{0}}}

\newcommand{\munull}{\ensuremath{\mu_{0}}}

\newcommand{\tauint}{\ensuremath{\tau_{\text{int}}}}

\newcommand{\kcr}{\ensuremath{\kappa_{\text{c}}}}
\newcommand{\kth}{\ensuremath{\kappa_{\text{t}}}}

\newcommand{\mps}{\ensuremath{m_{\text{PS}}}}
\newcommand{\bcusp}{\ensuremath{\beta_{\text{cusp}}}}
\newcommand{\bqu}{\ensuremath{\beta_{\text{qu}}}}

\newcommand{\ordpar}{\ensuremath{\langle \bar{\psi} i \gammafive\tauthree\psi\rangle}}
\newcommand{\ReL}{\ensuremath{\text{Re}(L)}}

\begin{document}

\begin{titlepage}
\begin{flushright}
DESY-09-065\\
HU-EP-09/20\\
MS-TP-09-06\\
SFB/CPP-09-41
\end{flushright}
\begin{centering}
\vfill
                   
{\bf\Large Phase structure of thermal 
lattice QCD with $N_f=2$ twisted mass Wilson fermions}

\vspace{0.8cm}

  E.-M. Ilgenfritz$^{1,2}$, K. Jansen$^3$, M. P. Lombardo$^4$, 
  M. M{\"u}ller-Preussker$^1$,\\
   M. Petschlies$^1$,  O. Philipsen$^5$, L. Zeidlewicz$^5$ 
 \vspace*{0.3cm}
  
  {\em
    $^1$ Institut f{\"u}r Physik, Humboldt-Universit{\"a}t zu Berlin, D-12489 Berlin, Germany\\
    $^2$ Institut f\"ur Theoretische Physik, Universit\"at Heidelberg,\\ 
         D-69120 Heidelberg, Germany\\
    $^3$ DESY, Zeuthen, D-15738 Zeuthen, Germany\\
    $^4$ Laboratori Nazionali di Frascati, INFN, I-100044 Frascati (RM), Italy\\
    $^5$ Institut f{\"u}r
    Theoretische Physik, 
    Westf{\"a}lische Wilhelms-Universit{\"a}t M{\"u}nster, D-48149 M\"unster, Germany
  }

%\vspace*{0.7cm}

\vspace*{0.7cm}
 
\begin{abstract}
\noindent
  We present numerical results for the phase diagram
  of lattice QCD at finite temperature in the formulation with 
  twisted mass Wilson fermions and a tree-level 
  Symanzik-improved gauge action. 
  Our simulations are performed on lattices with temporal 
  extent $N_{\tau}=8$, and lattice coupling $\beta$ ranging from strong
  coupling to the scaling domain.
  Covering a wide range in the space spanned by the lattice coupling $\beta$ and the hopping 
  and twisted mass parameters $\kappa$ and $\mu$, respectively, 
  we obtain a comprehensive picture of 
  the rich phase structure of the lattice theory.
  In particular, we verify the existence of an
  Aoki phase in the strong coupling region and the realisation of the 
  Sharpe-Singleton scenario at intermediate
  couplings. 
  In the weak coupling region we identify the phase boundary for the physical 
  finite temperature
  phase transition/crossover.
  Its shape in the three-dimensional parameter
  space is consistent with Creutz's conjecture of a cone-shaped thermal  
  transition surface. 
\end{abstract}
\end{centering}
 
\vfill

\end{titlepage}
 
\section{Introduction}
\label{sec:intro}

The quark-hadron transition predicted by Quantum Chromodynamics (QCD)
is a central aspect in describing strongly interacting matter at 
finite temperature $T$.
The early universe has passed through this transition, which is
now hoped to be reproduced in ultrarelativistic heavy ion collision experiments.
With a transition temperature 
$T_c\sim 150-200\,\text{MeV}$, the relevant energy scale is in the inherently 
non-perturbative regime of QCD, calling for numerical simulations 
of lattice QCD. We consider net baryon density zero, for which Monte Carlo
methods are directly applicable (cf. \cite{Philipsen:2007rj, DeTar:2008qi} 
for a review).
Most previous investigations of lattice thermodynamics have 
used the Kogut-Susskind (KS) staggered 
fermion formulation or the Wilson fermion formulation, 
amended by the improvement programme originally laid out by Symanzik 
\cite{Symanzik:1983dc, Symanzik:1983gh, Luscher:1984xn, Sheikholeslami:1985ij}. 
Both approaches have led to predictions for the nature of the transition, 
the transition temperature and the equation of state 
\cite{Ukita:2006pc, Bornyakov:2007zu, Aoki:2006br, Fodor:2007sy, Karsch:2007dp, Karsch:2007dt}.
In particular, based on simulations of a staggered action 
ref.~\cite{Fodor:2007sy} states that the finite 
temperature transition in continuum QCD at the physical point is analytic
with a transition temperature depending on the observable it is extracted from.
Given these results on the one hand and the continuing controversy 
on the universality of the staggered formulation 
\cite{Creutz:2007rk, Kronfeld:2007ek} on the other, a check of these findings
with fermions of the Wilson type appears desirable - 
even though the phase structure of the latter
is more complex compared to KS fermions due to explicit chiral symmetry breaking 
at $\mathcal{O}(a)$.

An economic alternative to the computationally involved Symanzik improvement  
and/or smearing techniques for the Wilson fermion formulation
might be Wilson twisted mass (Wtm) fermions 
\cite{Frezzotti:2000nk, Frezzotti:2003ni}. This formulation allows for automatic 
$\mathcal{O}(a)-$improvement of 
physically relevant operators by merely
tuning Wilson's hopping parameter $\kappa$ to its critical value (so called
maximal twist, cf. \cite{Shindler:2007vp} for a detailed review), without 
computational overhead. 
This fermion discretisation has been studied
extensively at zero temperature
by the European Twisted Mass Collaboration (ETMC)
\cite{Farchioni:2004us,Farchioni:2004ma,Farchioni:2004fs,Farchioni:2005ec,
Farchioni:2005bh,Farchioni:2005tu}
providing a growing amount of data relevant for tuning to maximal twist,
renormalisation factors and setting the scale 
\cite{Urbach:2007rt,Boucaud:2007uk,Boucaud:2008xu,Dimopoulos:2007fn}. 
In particular, simulations
at pseudoscalar masses $\mps \approx 300\,\text{MeV}$ are now feasible, and
considerable progress in controlling cut-off and finite size effects has been achieved by 
using lattice chiral perturbation theory \cite{Dimopoulos:2008sy}.
The price to pay for automatic improvement 
is a more complex phase structure of the lattice model 
due to the introduction of an additional parameter, the twisted mass $\mu_0$, as well as
explicit flavour symmetry breaking by the twisted mass term for $\mu_0\ne 0$.

The aim of our present study is to investigate the applicability 
of Wtm fermions for lattice thermodynamics with two degenerate 
quark flavours. For intermediate quark masses only a smooth crossover is expected in 
this case, which should turn into either a second or first order phase transition 
as the chiral limit is approached. The latter situation is, of course, very 
difficult to establish and currently still under debate, for some of the most recent attempts
with opposing conclusions, see \cite{Kogut:2006gt,Bonati:2009yi}.  
In this work, we restrict ourselves to a study of the phase structure of 
the Wtm fermion model with a  temporal lattice size $N_{\tau}=8$.
This is a prerequisite 
to establish whether simulations in the physically interesting regime and
using automatic $\mathcal{O}(a)$ improvement are feasible without running into unphysical phases.
The results we present here answer this question in the affirmative.

In Sec.~\ref{sec:TechnicalDetails} we specify the action and observables used in 
our simulations. The current knowledge about the phase structure  
for Wilson type fermions at finite temperature and that 
of Wtm fermions at zero temperature is summarised in Sec.~\ref{sec:Review}. 
We present our simulation results in Sec.~\ref{sec:SimRes} and discuss
them in the light of lattice chiral perturbation theory. 
Finally, Sec.~\ref{sec:Discussion} contains a discussion and our conclusions.
Preliminary results of this work have already been
reported in \cite{Ilgenfritz:2006tz, Ilgenfritz:2007qr, Ilgenfritz:2008td}.

\section{Action and observables}
\label{sec:TechnicalDetails}

Our setup consists of a hypercubic lattice $N_{\sigma}^3\times N_{\tau}$
with periodic boundary conditions for the gauge fields,
whereas the fermionic fields are subject to (anti-) periodic boundary
conditions in the (time) space direction.
We use the tree-level Symanzik improved gauge action given by
\begin{equation}
  \Act_{g}[U] = \beta\left[ 
    c_0\,\sum\limits_{P}\left(1-\frac{1}{3} \text{Re} (\Trace[U(P)]) \right) +
    c_1\,\sum\limits_{R}\left(1-\frac{1}{3} \text{Re} (\Trace[U(R)]) \right) 
  \right]\;,
  \label{equ:tlWilsonGaugeAction}
\end{equation}
where $\beta=6/g_0^2$ with $g_0$ the bare gauge coupling and $U(P), U(R)$ 
denote the path-ordered products of link variables along closed loops of 
length $4a$ (plaquettes) and $6a$ (rectangles), respectively.
The weight coefficients satisfy $c_0 + 8\,c_1=1$ (normalisation)
and $c_1 = -1/12$ (tree-level improvement condition). This choice of gauge action is 
motivated by allowing us to exploit ETMC's data for $T=0$.

The Wilson twisted mass action for the fermion sector in the so-called twisted basis reads
\begin{equation}
  \Act_F[\psi\,,\psib\,,U] = a^4\sum\limits_{x}
  \left[
     \psib(x) \left( D_W + \mnull + i\munull\gammafive\tauthree\right) \psi(x)
  \right]\;,
  \label{equ:WtmAction}
\end{equation}
with $D_W = \gammamu\,\frac{1}{2a}\left(\nablamu + \nablamu^*\right) -\frac{ar}{2}\nablamu\nablamu^*$
the Wilson-Dirac operator. This action can be rewritten as
\begin{equation}
  \begin{split}
    \Act_F[\Psi\,,\Psib\,,U] &= a^4\sum\limits_{x}
    \Big[
       \Psib(x) \big( \gammamu\,\frac{1}{2a}\left(\nablamu + \nablamu^*\right) 
       -\frac{ar}{2}{\rm e}^{-i\omega\gammafive\tauthree}  \nablamu^* \nablamu +M_0 \big) \Psi(x)
  \Big]
  \end{split}
  \label{equ:WtmActionPhysicalBasis}
\end{equation}
in the so-called physical basis $\Psib,\,\Psi$, which is related to the 
twisted basis $\psib,\,\psi$ by the non-anomalous chiral rotation 
\begin{align}
  \Psi  &= \exp\left(i\frac{\omega}{2}\gammafive\tauthree\right)\psi\;, \nn\\
  \Psib &= \psib \exp\left(i\frac{\omega}{2}\gammafive\tauthree\right)\;,
  \label{equ:Rotation}
\end{align}
with 
\begin{equation}
  M_0=\sqrt{\mnull^2 + \mu_0^2}\,, \qquad \tan(\omega)=\mu_0/\mnull
  \label{equ:def_M0}
\end{equation}
denoting the bare polar quark mass and the bare twist angle, respectively.
Finally, introducing Wilson's hopping parameter $\kappa = 1/(2a\mnull+8r)$ and rescaling 
the fermion fields according to $\psi \rightarrow \sqrt{a^3/(2\kappa)}\psi$ leads to the familiar
form
\begin{equation}
  \begin{split}
    \Act_F&[\psi\,,\psib\,,U] = \sum\limits_{x} 
    \Big[
    \psib(x) \left(1 + i2\kappa a\mu_0\gammafive\tauthree \right)\psi(x) \\
    & \quad -\kappa\sum\limits_{\mu}\psib(x)\left(
      (r-\gammamu)\Umux\psi(x+\hmu) + (r+\gammamu)\Umuxmad\psi(x-\hmu)
    \right)
    \Big]\;.
  \end{split}
  \label{equ:WtmAction2}
\end{equation}
Because of the spin-flavour structure of the twisted mass term, 
$i a\munull\gammafive\tauthree$, 
parity is only a symmetry when combined with a discrete flavour rotation or a sign change
$\mu_0\rightarrow -\mu_0$, while
flavour symmetry is broken explicitly according to the pattern 
$SU_V(2)\rightarrow U_{(3)}(1)$,
where $U_{(3)}(1)$ is the subgroup of flavour rotations generated by $\tauthree$.

For our numerical simulations we use a generalised hybrid Monte-Carlo algorithm with even-odd
preconditioning \cite{DeGrand:1988vx}, Hasenbusch trick 
\cite{Hasenbusch:2001ne, Hasenbusch:2003vg} 
and multiple time scale integration according to the Sexton-Weingarten scheme 
\cite{Weingarten:1991ra} (cf. \cite{Urbach:2005ji} for algorithmic benchmarks
and \cite{Jansen:2009xp} for a more recent collection of algorithmic improvements). 

In our investigation of the phase structure we measure and analyse the following
observables:
\begin{enumerate}
\item (averaged) plaquette $P$
  \begin{equation}
    P = \frac{1}{6N_{\sigma}^3N_{\tau}}\sum\limits_{x}\sum\limits_{\mu<\nu} 
    \Trace\left[
      U_{\mu}(x)U_{\nu}(x+\hmu)U_{\mu}(x+\hnu)^{\dagger}U_{\nu}(x)^{\dagger}
    \right]\;,
  \end{equation}
\item (real and imaginary part of the) Polyakov loop $L$
  \begin{equation}
    L = \frac{1}{3N_{\sigma}^3}\sum\limits_{\vec{x}} 
    \Trace\left[\prod\limits_{i=0}^{N_{\tau}-1}U_{4}(\vec{x},i) \right]\;,
  \end{equation}
\item their susceptibilities according to
  \begin{equation}
    \chi(\mathcal{O}) = N_{\sigma}^3 \left( \langle \mathcal{O}^2 \rangle - 
      \langle \mathcal{O}\rangle^2 \right)\;,
  \end{equation}
\item fermionic observables: 
  scalar condensate $\langle \psib \psi \rangle$, the order parameter for
  parity-flavour symmetry breaking
  $\langle \psib i\gammafive\tauthree\psi \rangle$ and the pion norm
  \begin{equation}
    |\pi|^2 = \sum\limits_{x}\langle \psib(x)\gammafive\frac{\tau^+}{2}\psi(x)
    \psib(0)\gammafive\frac{\tau^-}{2}\psi(0) \rangle\;,
  \end{equation}
  where $\tau^{\pm} = \tauone \pm i\tautwo$.
\end{enumerate}
Discontinuous phase transitions are in principle indicated by any observable, 
but depending on the particular dynamics one may be more sensitive than the other.
In particular, the Polyakov loop is an order parameter for $Z(N)$-breaking and hence
susceptible to signals of deconfinement, which get more pronounced
the heavier the quark mass. On the other hand, the pion norm and scalar condensate 
are sensitive to small eigenvalues of the Wtm operator and hence detect signals
of chiral symmetry breaking in the light mass regime. 
The pseudo-scalar condensate is the order parameter for 
spontaneous parity-flavour symmetry breaking in the plane 
$\{\mu_0=0\}$ of the $(\beta,\,\kappa,\,\mu_0)-$parameter
space \cite{Aoki:1983qi, Aoki:1997fm, Ilgenfritz:2003gw}.

\section{Current knowledge of the phase structure}
\label{sec:Review}

\subsection{Wilson fermions at $T\neq 0$}
\label{subsec:WilsonFermionPhaseStructure}

\begin{figure}[t]
  \begin{center}
    \begin{minipage}[ht]{0.495\textwidth}
      \includegraphics[width=0.7\textwidth,angle=-90]{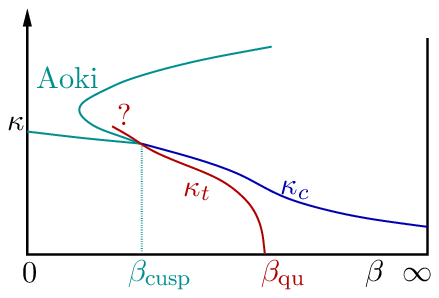}
    \end{minipage}
    \begin{minipage}[ht]{0.495\textwidth}
      \includegraphics[width=0.9\textwidth]{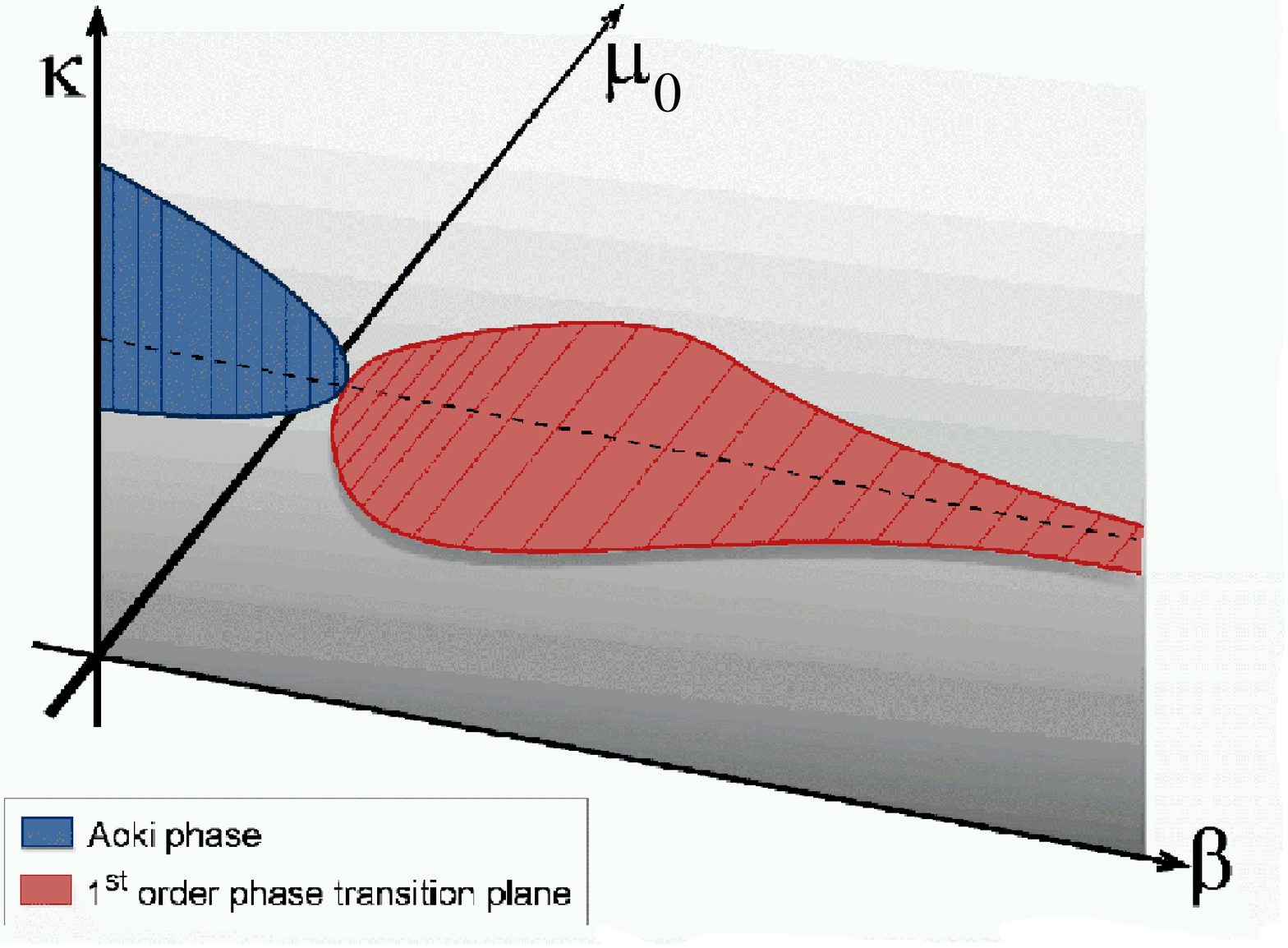}
    \end{minipage}
    \caption[]{Left: Aoki phase and thermal transition $\kappa_t(\beta)$ 
                   for untwisted Wilson fermions. 
                   $\beta_{\rm cusp}$ denotes the tip of the Aoki phase
                   and $\beta_{\rm qu}$ the critical coupling for the deconfinement 
                   transition in the quenched limit.
             Right: Aoki phase and first order bulk transition for twisted mass Wilson
                    fermions at zero temperature \cite{Shindler:2007vp}.
      \label{fig:aoki}}
  \end{center}
\end{figure}

The phase structure for standard and improved Wilson fermions is 
determined by two phenomena: spontaneous parity-flavour symmetry breaking
and the finite temperature transition. The existence of a phase
of spontaneously broken parity-flavour symmetry, commonly called Aoki phase, 
was introduced theoretically and verified numerically in 
\cite{Aoki:1983qi, Aoki:1997fm, Aoki:1986ua, Aoki:1987us, Aoki:1995yf}. 
Refs.~\cite{Ilgenfritz:2003gw,Bitar:1996kc,Sternbeck:2003gy, Ilgenfritz:2005ba} 
demonstrated that the Aoki phase is 
located in the strong coupling region and ends in a cusp 
at a finite $\bcusp$, from which the chiral critical line $\kcr(\beta)$ 
emanates and continues towards the weak coupling limit, cf.~Fig.~\ref{fig:aoki} (left).
The Aoki phase is repeated for larger values of $\kappa$, corresponding to the fermion 
doubling phenomenon. The value of
$\bcusp$ depends on the choice of action and the temporal lattice 
extent.  Being a lattice artefact, the Aoki phase does not affect physics in 
the weak coupling region. 

The finite temperature transition or crossover 
is characterised by a physical transition temperature which depends on the renormalised 
quark mass, $T_c(m_R)$, and therefore, through $m_R$, on the bare quark mass. 
Since $T= 1/(a(\beta)N_{\tau})$ on the lattice and the bare quark mass is 
parametrised in terms of $\kappa$, pairs of
$(m_R,\,T_c)$ translate to pairs $(\kappa_t(\beta_c),\,\beta_c)$ on the lattice describing
the phase boundary for finite temperature transitions at fixed $N_{\tau}$.
This line has been mapped out in detail by  
the CP-PACS \cite{Ukita:2006pc,AliKhan:2000iz, AliKhan:2001ek} and
DIK \cite{Bornyakov:2007zu,Bornyakov:2004ii, Bornyakov:2005dt} collaborations
for the case of an $\mathcal{O}(a)$-Symanzik-improved fermion
action on one hand and the standard plaquette or the renormalisation group 
improved gauge action on the other with the following results:
\begin{enumerate}
\item The thermal line  for fixed $N_{\tau}$  
  runs from a finite $\bqu$ in the quenched limit 
  ($\kth=0$), where the transition is of first order,
  towards the zero temperature chiral critical 
  line $\kcr(\beta,T=0)$ and lower $\beta$.
  This is consistent with the physical expectation that
  the transition temperature increases with the quark mass.

\item In the vicinity of $\kcr(\beta,T=0)$ and the cusp of the Aoki phase,
  the thermal line appears to run close and parallel to the critical line; 
  despite large computational efforts a clear separation of the thermal line 
  from the border line of the Aoki phase has not yet been achieved on 
  $N_{\tau}=4$ lattices \cite{Ilgenfritz:2005ba} 
  in case of the pure Wilson gauge action.
\end{enumerate}

\subsection{Twisted mass fermions at $T=0$}
\label{subsec:WtmFermionPhaseStructure}

The zero temperature phase structure of Wilson twisted mass fermions has 
been investigated in a series of publications by the ETM Collaboration 
\cite{Farchioni:2004us,Farchioni:2004ma,Farchioni:2004fs,Farchioni:2005ec,
Farchioni:2005bh,Farchioni:2005tu}.
The parameter space is extended from the $\{\beta,\kappa\}-$plane to the 
3d $\{\beta,\kappa,\mu_0\}-$space. Hence, in the strong coupling region 
an Aoki phase should exist in the $\beta-\kappa-$plane for 
$0\le\beta\le\bcusp$, cf.~Fig.~\ref{fig:aoki} (right). 
Adjacent to it a surface of first order phase transitions extends into 
the weak coupling region. It is perpendicular to the Aoki phase and includes the
chiral transition line at zero twist. 
This surface is predicted from chiral perturbation theory \cite{Sharpe:1998xm, Munster:2004am} 
and has been established in zero temperature simulations \cite{Farchioni:2004us}.
This transition for $\munull>0$ is also a lattice artefact, 
correspondingly its width in the $\munull-$direction 
decreases with increasing $\beta$ and can be made smaller at a fixed value of 
$\beta$ by using improved gauge actions \cite{Farchioni:2004fs}. Finite temperatures
correspond to finite $N_\tau$ on the lattice, which will amount to small shifts
of the transition surface compared to $T=0$.

These unphysical first order transitions and their suppression are a crucial issue
for simulations at small quark and corresponding pseudoscalar masses. 
At maximal twist \cite{Boucaud:2007uk, Boucaud:2008xu}, realised by setting
$\kappa=\kcr(\beta)$, 
the width of the surface in $\mu_0$-direction implies a lower bound 
$\mu_{0\text{c}}(\beta)$ for the twisted mass and hence for the quark mass 
(cf.~Eq.~(\ref{equ:def_M0})) at fixed $\beta$, in order to avoid unphysical 
phase transitions.

\subsection{Twisted mass fermions at $T\neq 0$: Creutz's cone conjecture}
\label{subsec:CCC}

In \cite{Creutz:1996bg,Creutz:2007fe} Creutz conjectured a finite
temperature phase structure for untwisted and twisted Wilson fermions based on 
continuum symmetry arguments. 

\begin{figure}[t]
  \vspace*{-1.0cm}
  \begin{center}
    \includegraphics[width=0.6\textwidth,angle=-90]{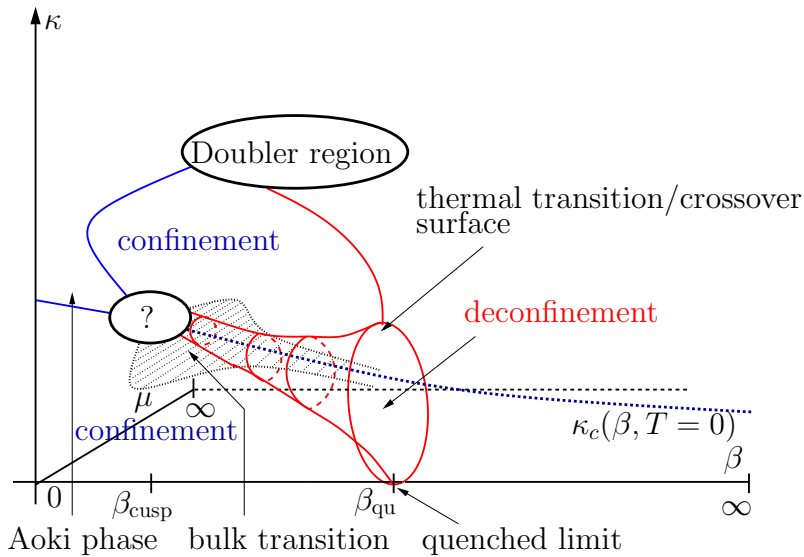}
    \vspace*{-1.5cm}
    \caption{Schematic view of Creutz's cone conjecture, derived from
      \cite{Creutz:2007fe}.
      \label{fig:CCC}
    }
  \end{center}
  \vspace*{-0.5cm}
\end{figure}

In the continuum theory the twist angle $\tan(\omega)=\munull/\mnull$ 
is an irrelevant parameter and
the rotation $\exp(i\omega/2\gammafive\tauthree)$ a mere change of integration variables
in the path integral, leaving the path integral measure invariant. Hence, $T_c$ can only 
be a function of the renormalised quark mass $M_R = \sqrt{m_R^2+\mu_R^2}$, 
i.\,e.~$T_c$ is constant on lines of constant $M_R$, which correspond to circles.
In bare parameter space, because of the renormalisation factors   
for the masses these curves are ellipses.
This picture should translate to the lattice, modified by cut-off effects.
For fixed $\beta$, the lines of constant $M_R$
are closed curves that become ellipses in the continuum limit. 
Hence, for fixed $N_{\tau}$ there
should be an interval  $\bcusp \lesssim\beta \lesssim \bqu$ where the surface 
of finite $T$ transitions wraps around the chiral transition line.
Since $T_c$ grows with $M_R$, this surface should have a
conical shape opening in the direction of growing $\beta$.
Fig.~\ref{fig:CCC} gives a schematic summary of the expected phase structure. 

While the existence of a finite temperature transition for every value of the 
twist angle is dictated by the continuum theory,
the strength and nature of the transition are subject to discretisation effects and thus generically 
dependent on the twist angle $\omega$. This will lead to a distortion of the 
ellipse and conical shape. For sufficiently light quarks, 
this expectation can be made more quantitative in the 
framework of lattice chiral perturbation theory. At next-to-leading order (NLO), 
the expression for the pion mass is \cite{Sharpe:2004ny,Sharpe:2006pu}
\begin{gather}
m_{\pi^\pm}^2 = M' + \frac{16}{f^2}\left( (2L_{68}-L_{45})(M')^2 + 
M' \hat{a} \cos(\omega) (2W-\tilde{W})+2\hat{a}^2\cos^2(\omega)W'\right)
\label{eq:mpi}\\ + 
\frac{(M')^2}{2\Lambda_\chi^2}\ln\left(\frac{M'}{\Lambda_R}\right)\;,\nonumber
\end{gather}
where the quark masses are now renormalised ones, 
$\mu =Z_\mu \mu_0$,
  $m =Z_m (m_0-m_c)$.
Furthermore, $M'=\sqrt{ \hat{\mu}^2+ (\hat{m}')^2}$, $\hat{a}=2W_0a$, and 
$\hat{\mu}=2B_0Z_\mu\mu_0$, 
$\hat{m}'=2B_0Z_m(m_0-m_c)$.
The $B$'s, $W$'s, $L$'s can in principle
all be fixed by fitting lattice chiral perturbation theory to 
zero temperature simulation results. Currently, this is not completed yet.
Once the NLO formula describes the data sufficiently well, knowledge of
one thermal transition point on the cone will suffice to predict the entire
distorted ellipse for a given $\beta$ and pion mass.

\section{Simulation results}
\label{sec:SimRes}

The goal of our numerical simulations is to establish and locate 
the different phases discussed in the previous section. 
The following analysis of our numerical data thus proceeds along the 
$\beta-$axis from strong coupling ($\beta \lesssim 1.8$) to the 
scaling region ($\beta \gtrsim 3.8$).
The parameter sets for which simulations have been carried out are 
quoted in Tab. \ref{tab:SimulationRuns} in the Appendix. 
Throughout this paper we have used lattices with temporal size $N_\tau=8$
and spatial size $N_{\sigma}=16$, except in a
few cases where it will be indicated explicitly.

\begin{figure}[t]
        \includegraphics[width=0.35\textwidth,angle=-90]{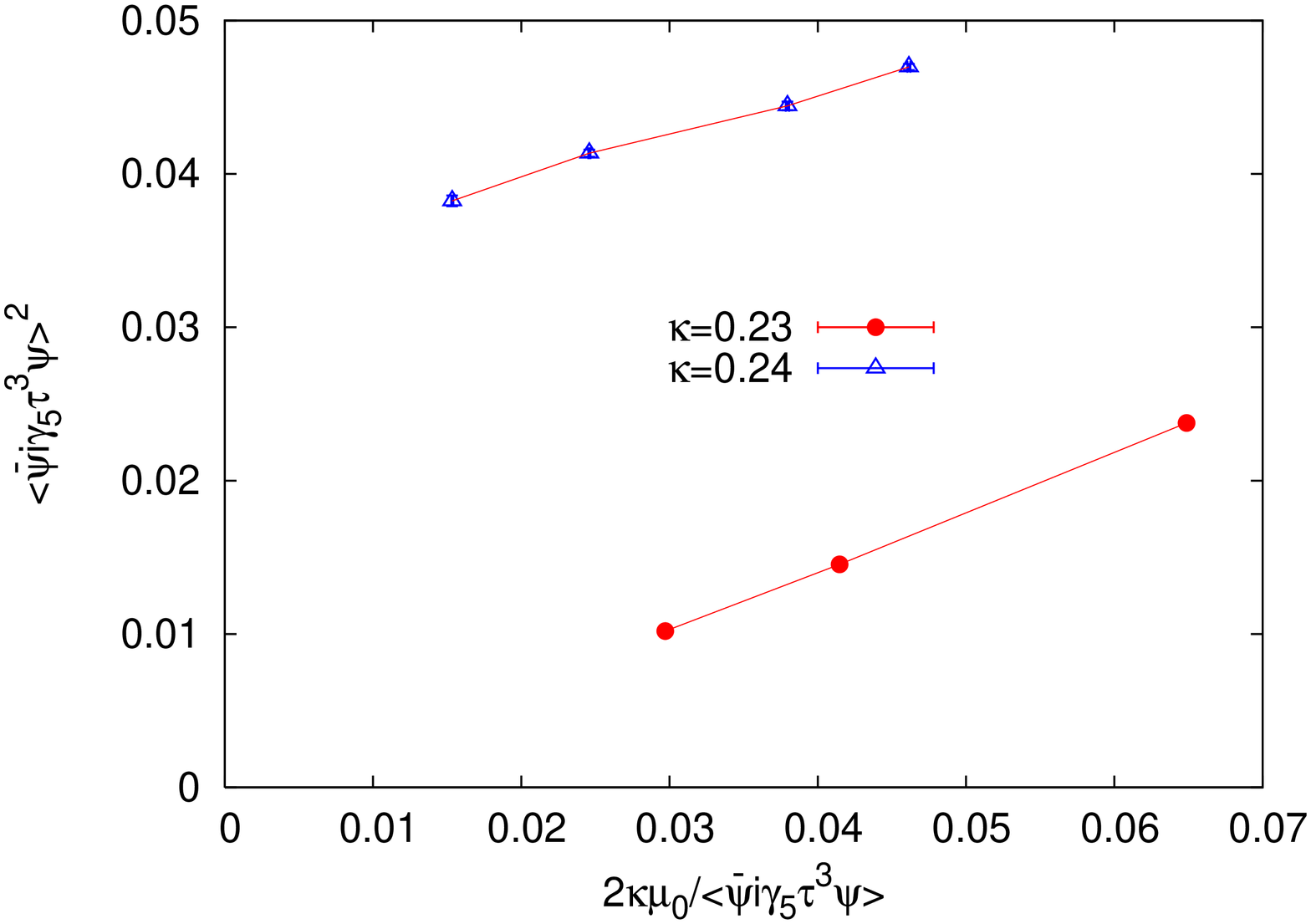}
        \includegraphics[width=0.35\textwidth,angle=-90]{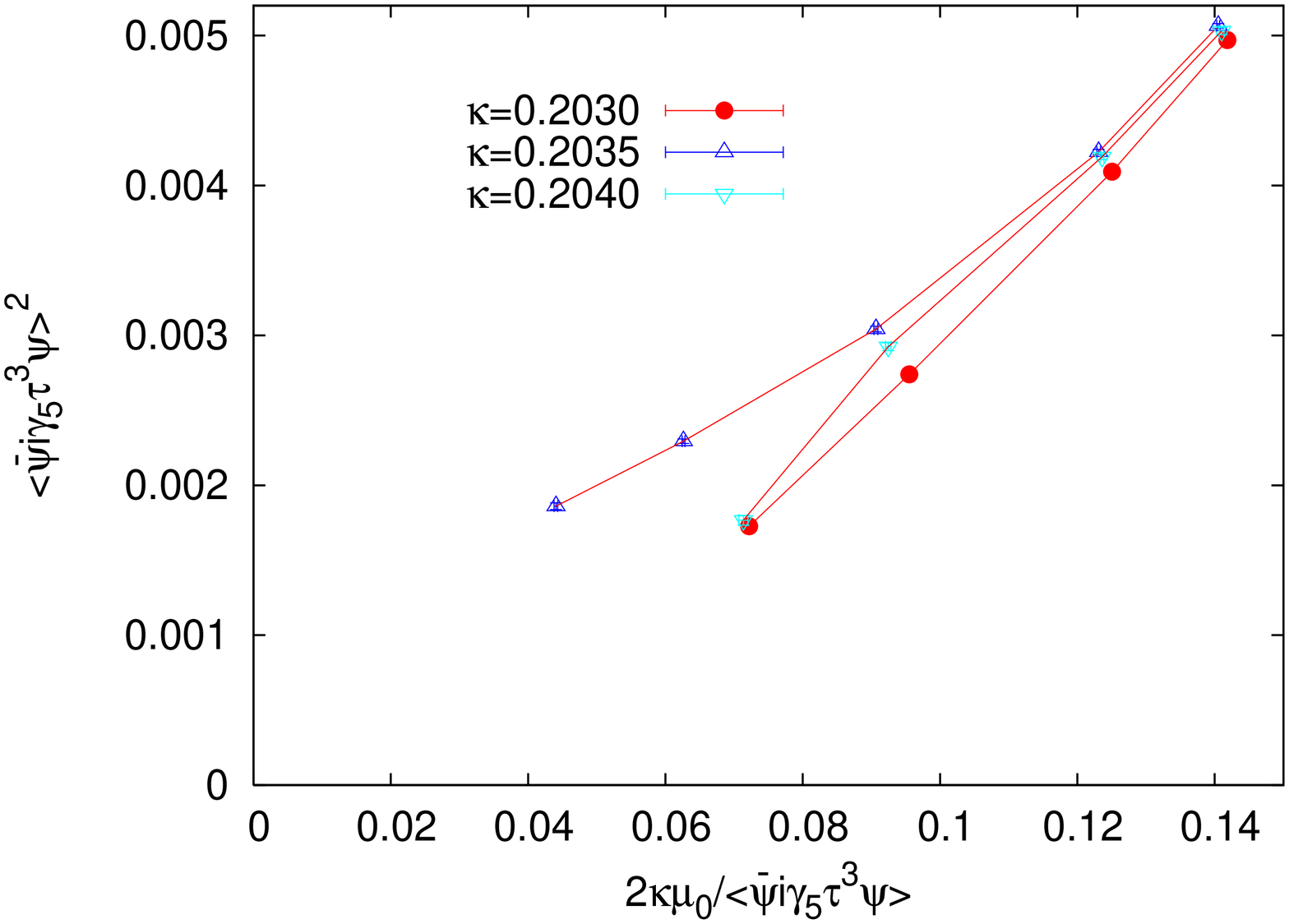}
        \includegraphics[width=0.35\textwidth,angle=-90]{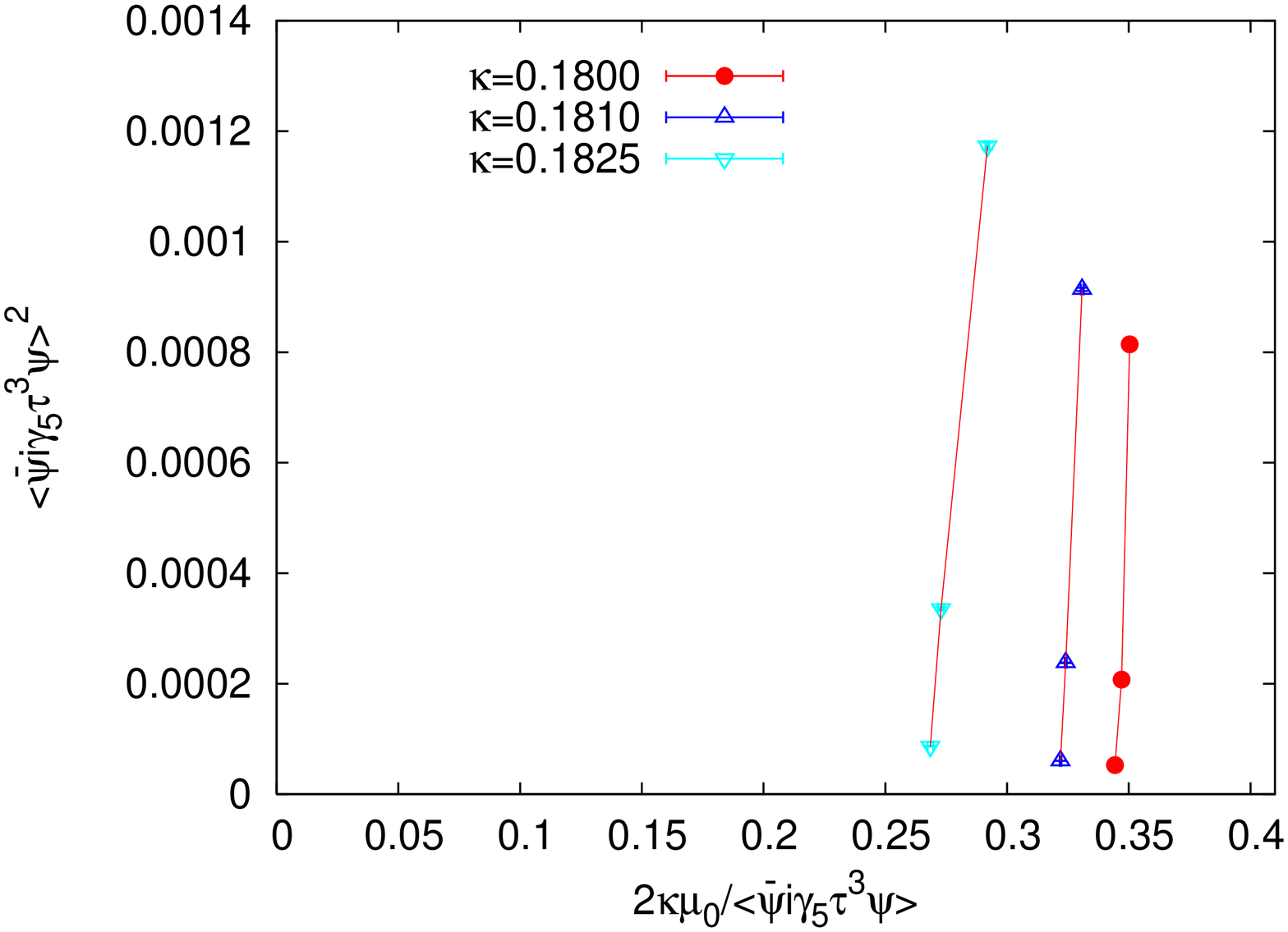}
        \includegraphics[width=0.35\textwidth,angle=-90]{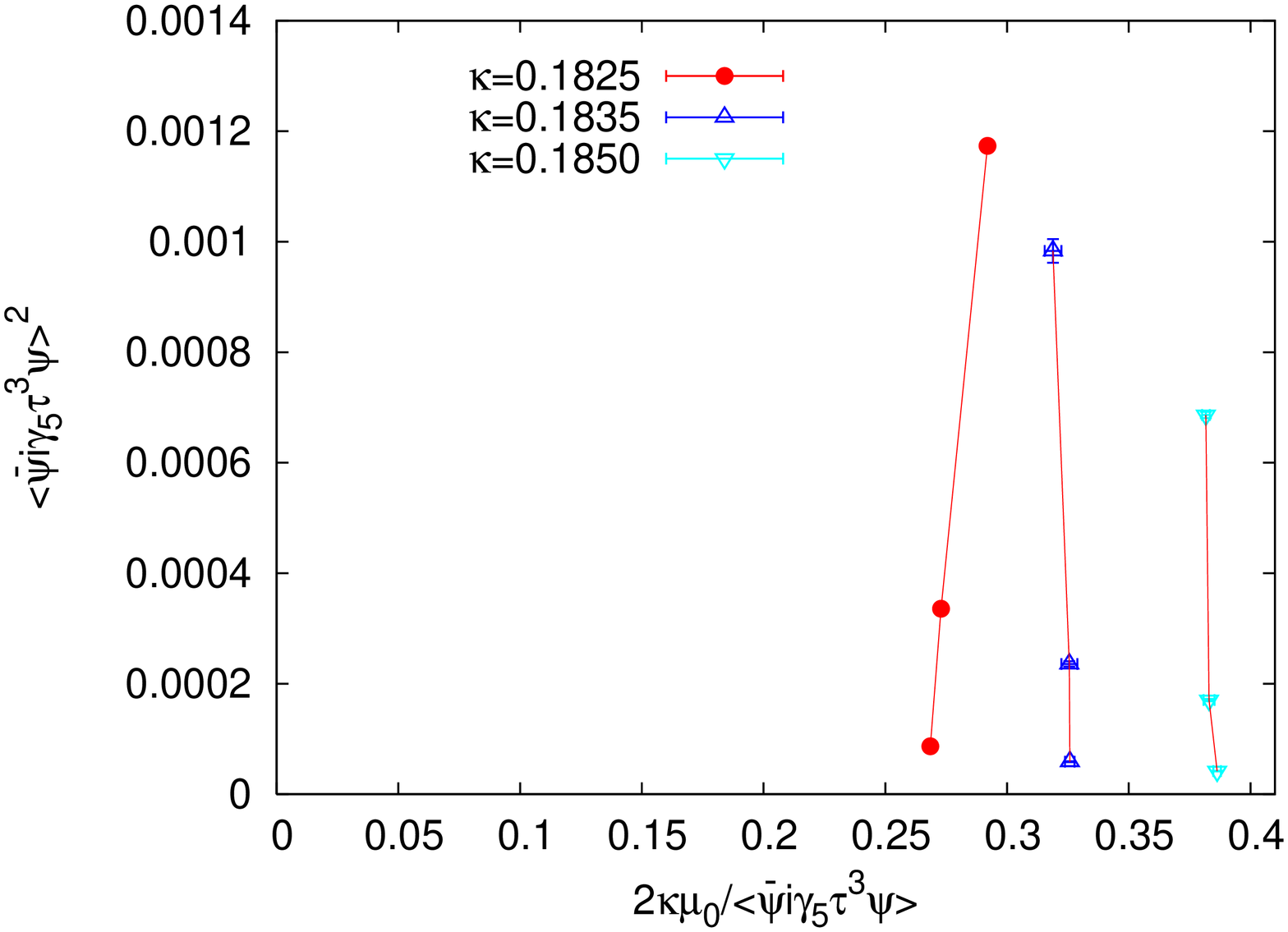}
    \caption[]
    {Fisher plots for $\beta=1.8$ (upper left panel), $\beta=3.0$  
    (upper right panel) and $\beta=3.4$ (lower panels) for 
    $\kappa<\kcr\approx 0.1825$ (left) and $\kappa>\kcr$ (right).
      \label{fig:AokiPhase}}
\end{figure}
\subsection{Strong coupling: the Aoki phase}
\label{subsec:AokiSearch}

We checked for the existence of an Aoki phase at $\beta=1.8,\,3.0,\,3.4$. 
For the analysis we consider so-called
Fisher plots showing $\ordpar^2$ as a function of $2\kappa\munull/\ordpar$  
(cf.~\cite{Ilgenfritz:2003gw} and references therein for a motivation of this 
approach). One then inspects the behaviour in the limit $2\kappa\munull/\ordpar \to 0$:
for $(\beta,\,\kappa)$ points outside the Aoki phase the curves will hit the 
abscissae at a non-vanishing value of $2\kappa\munull/\ordpar$, a 
point on the phase boundary would lead to a straight line through the origin and
a point within the Aoki phase would show a
positive intercept with the ordinate axis. Our results are shown in 
Fig.~\ref{fig:AokiPhase}.

The upper left panel shows the situation for the strongest coupling we considered,
$\beta=1.8$. We checked for finite-volume effects on lattices with 
$N_{\sigma}=16,\,24,\,32$ and  found them to be negligible within the statistical 
uncertainty. From the curves we conclude that 
the point $(\beta,\,\kappa)=(1.8,\,0.24)$ lies inside the
Aoki phase, whereas $(\beta,\,\kappa)=(1.8,\,0.23)$ is outside.
At $\beta=3.0$ the upper right panel suggests that the Aoki phase has narrowed to
the interval $0.203 < \kappa < 0.204$. The curve for $\kappa=0.2035$ was 
measured for $N_{\sigma}=16$ only. 
The point $(\beta,\,\kappa)=(3.0,\,0.2035)$ appears to be a candidate for 
spontaneous parity-flavour symmetry breaking.

\subsection{Intermediate coupling: the Sharpe-Singleton scenario}

\begin{figure}[t]
        \includegraphics[width=0.5\textwidth,angle=0]{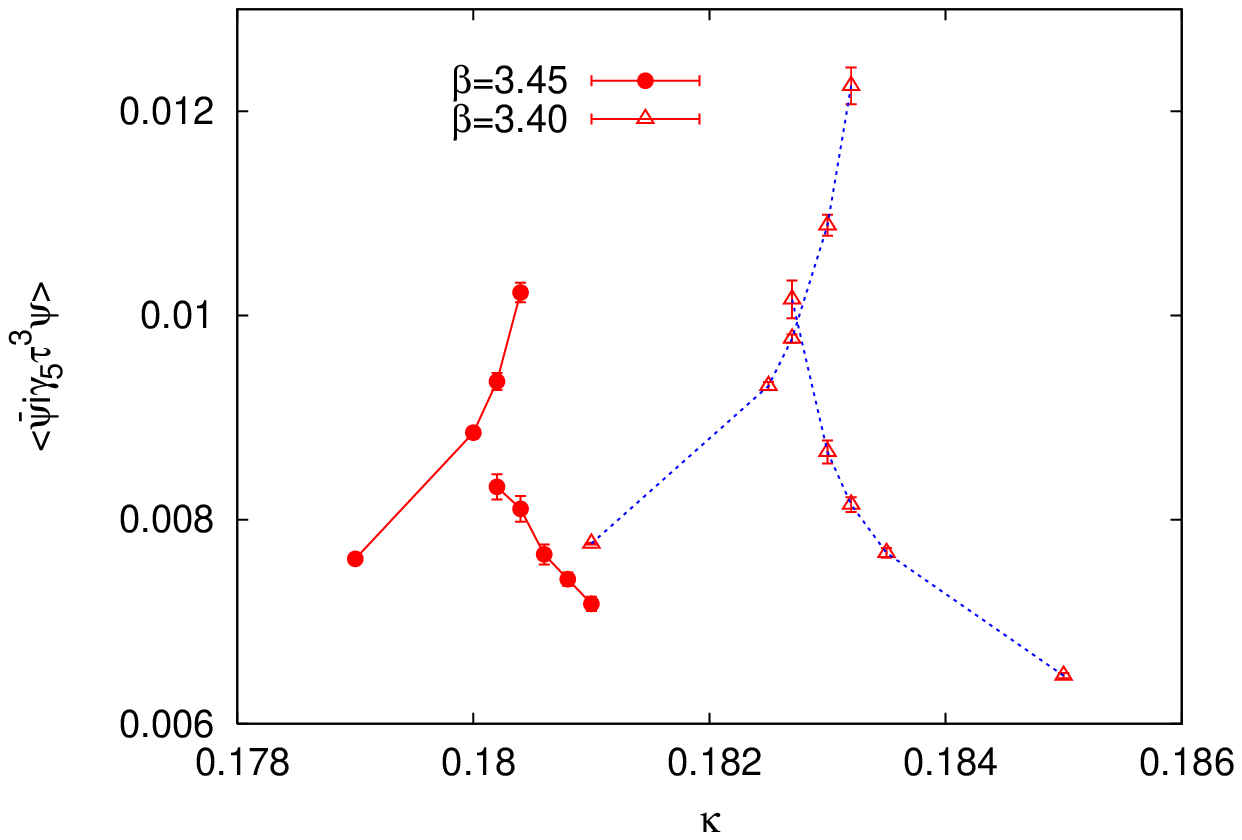}
        \includegraphics[width=0.5\textwidth,angle=0]{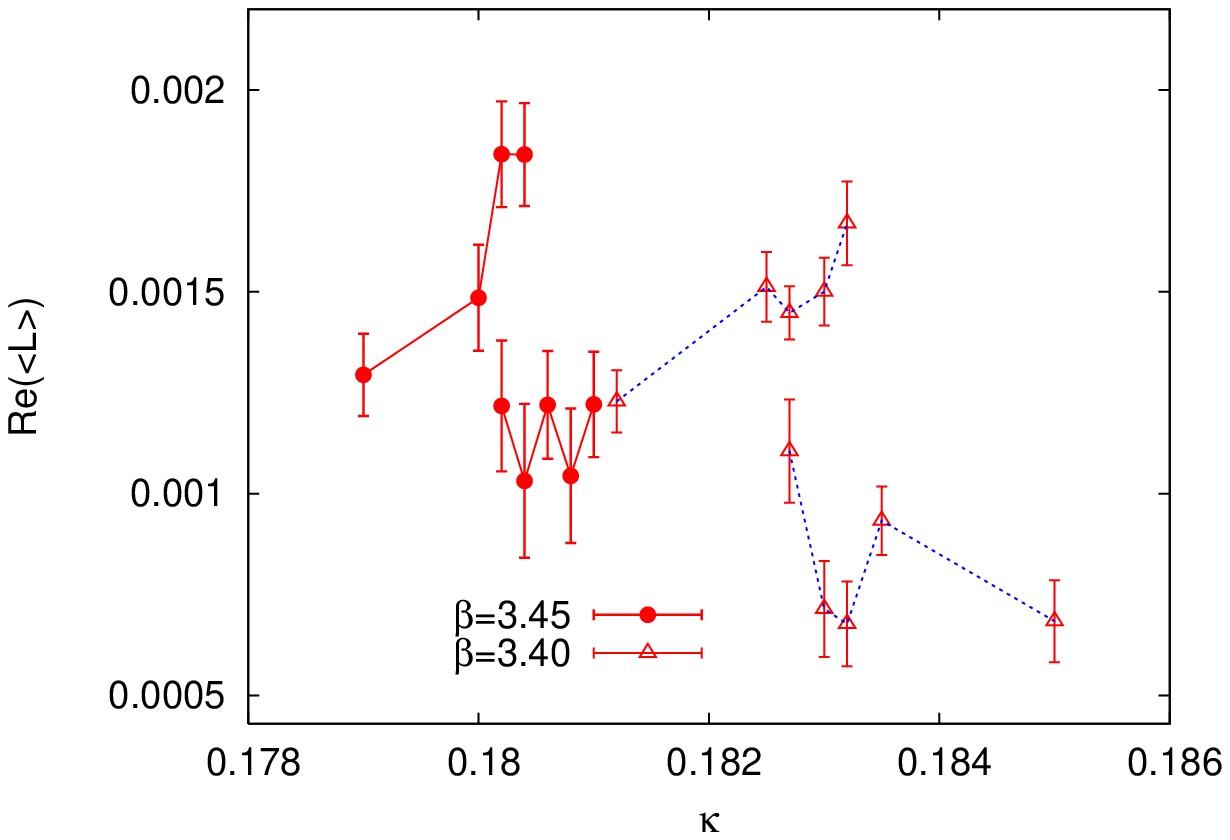}
    \caption[]
    {Metastability signals in the order parameter 
    $\langle \psib i\gammafive\tauthree\psi \rangle$ (left) and the Polyakov loop (right)
     in the neighbourhood of the critical line, $\munull\approx 0.0068$. 
     The lines were added for visual guidance.
      \label{fig:MetaStab}}
\end{figure}

The situation changes notably when proceeding to $\beta=3.4$,
as shown in the lower panels of Fig.~\ref{fig:AokiPhase}. Neither below nor above
$\kcr\approx 0.1825$ do we see any indication of
spontaneous symmetry breaking. Hence, within our resolution 
the existence of an Aoki phase is ruled out for $\beta=3.4$.
This assertion is supported by the
emergence of another phenomenon.
We find clear signals of metastability 
in all observables when scanning in  
$\kappa$ across the region of the chiral critical line,
provided $\munull$ was chosen small enough. 
Fig.~\ref{fig:MetaStab} shows this behaviour for two different observables.
We interpret these two-state signals as tied to the surface of 
first order phase transitions discussed in Sec.~\ref{subsec:WtmFermionPhaseStructure}.
(Note that in \cite{Blum:1994eh} 
metastability signals were also reported in simulations
with pure Wilson fermions at finite temperature and $N_\tau=6$).

\subsection{Weak coupling: finite temperature transition near $\kcr$}
\label{subsec:FiniteTatKcr}

\begin{figure}[t]
        \includegraphics[width=0.35\textwidth,angle=-90]{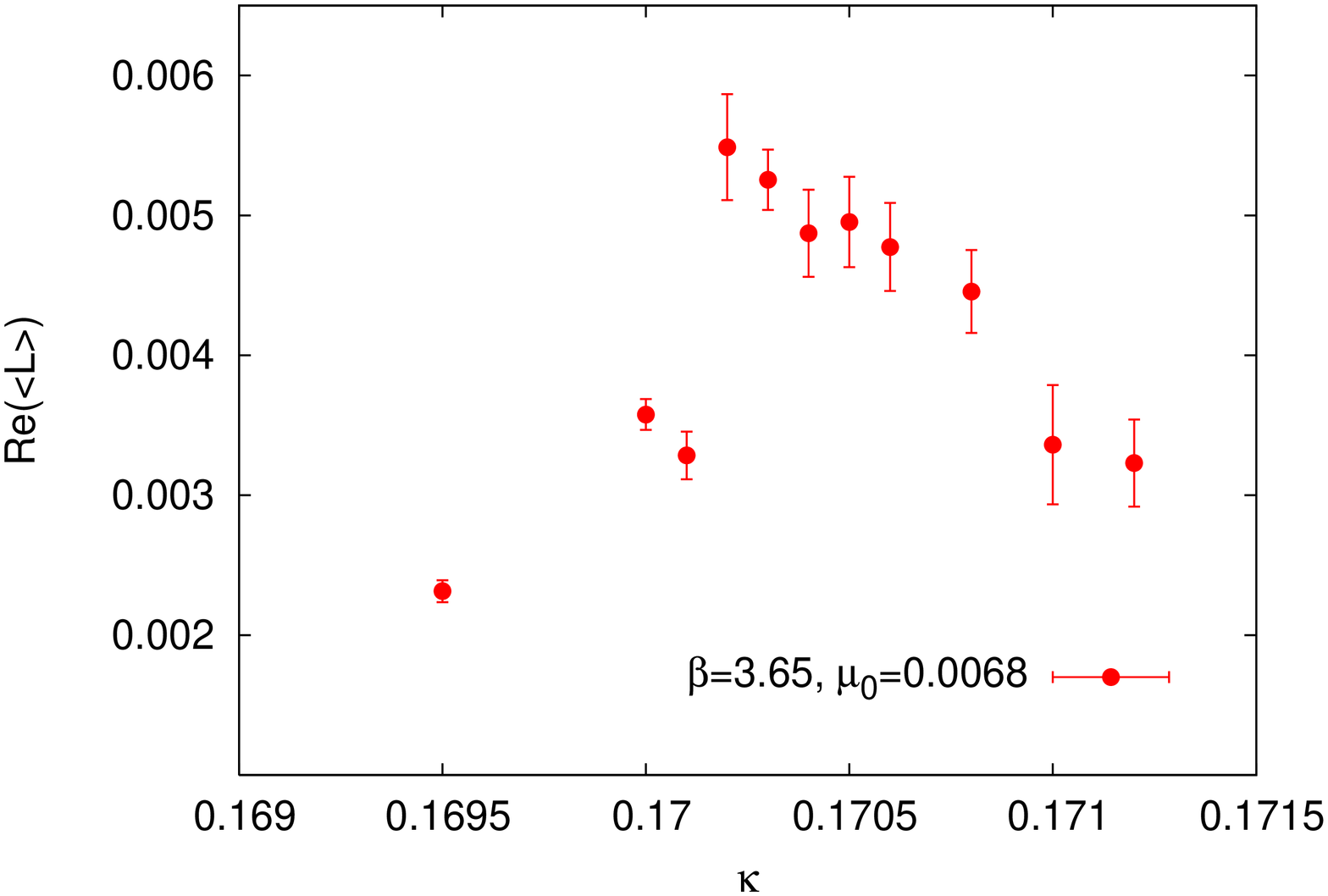}
        \includegraphics[width=0.35\textwidth,angle=-90]{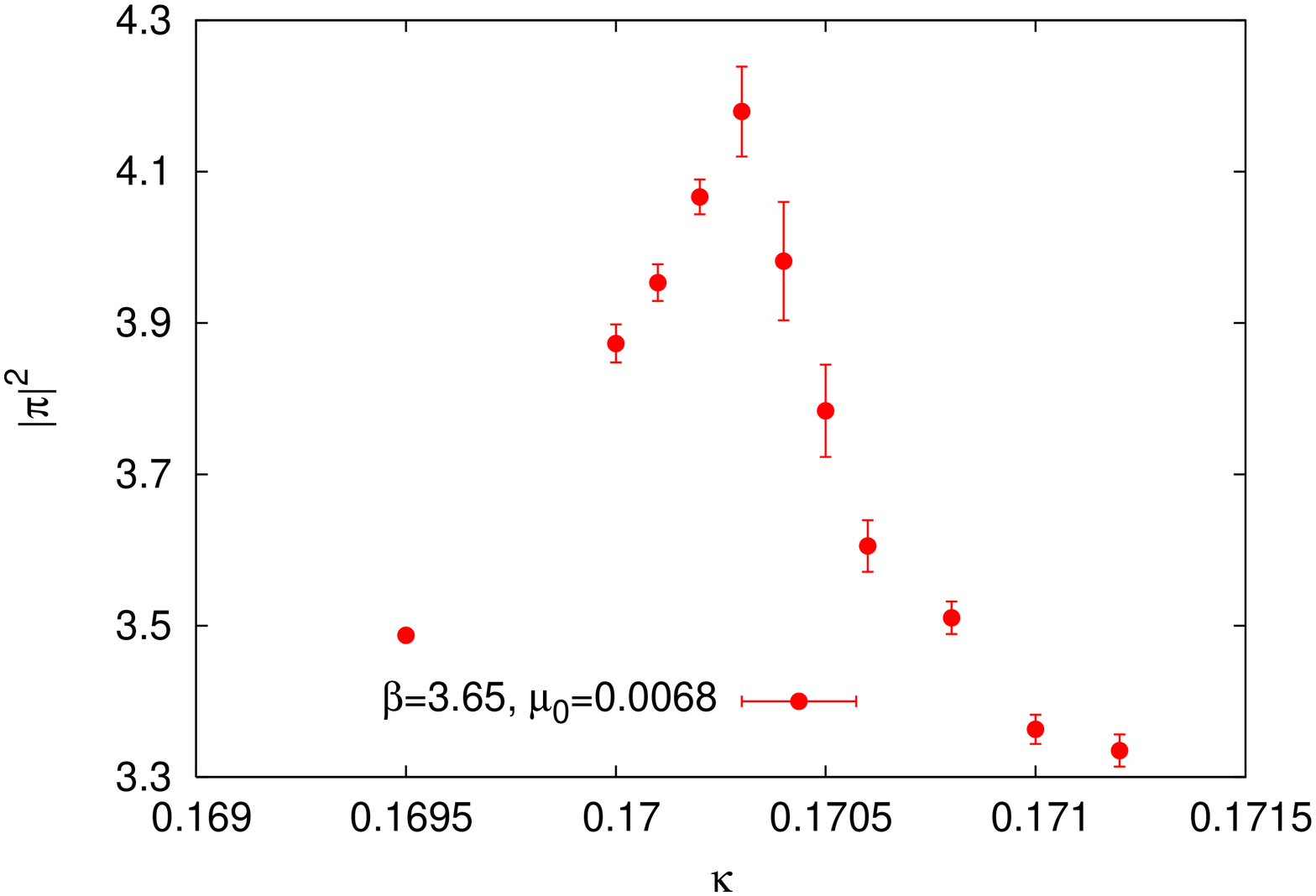}
    \caption[]{
      Real part of the Polaykov loop (left) and the pion norm (right) around $\kcr$
      at $\beta=3.65$ and $\munull=0.0068$.
      \label{fig:FiniteTb365}}
\end{figure}
\begin{figure}[t]
        \includegraphics[width=0.35\textwidth,angle=-90]{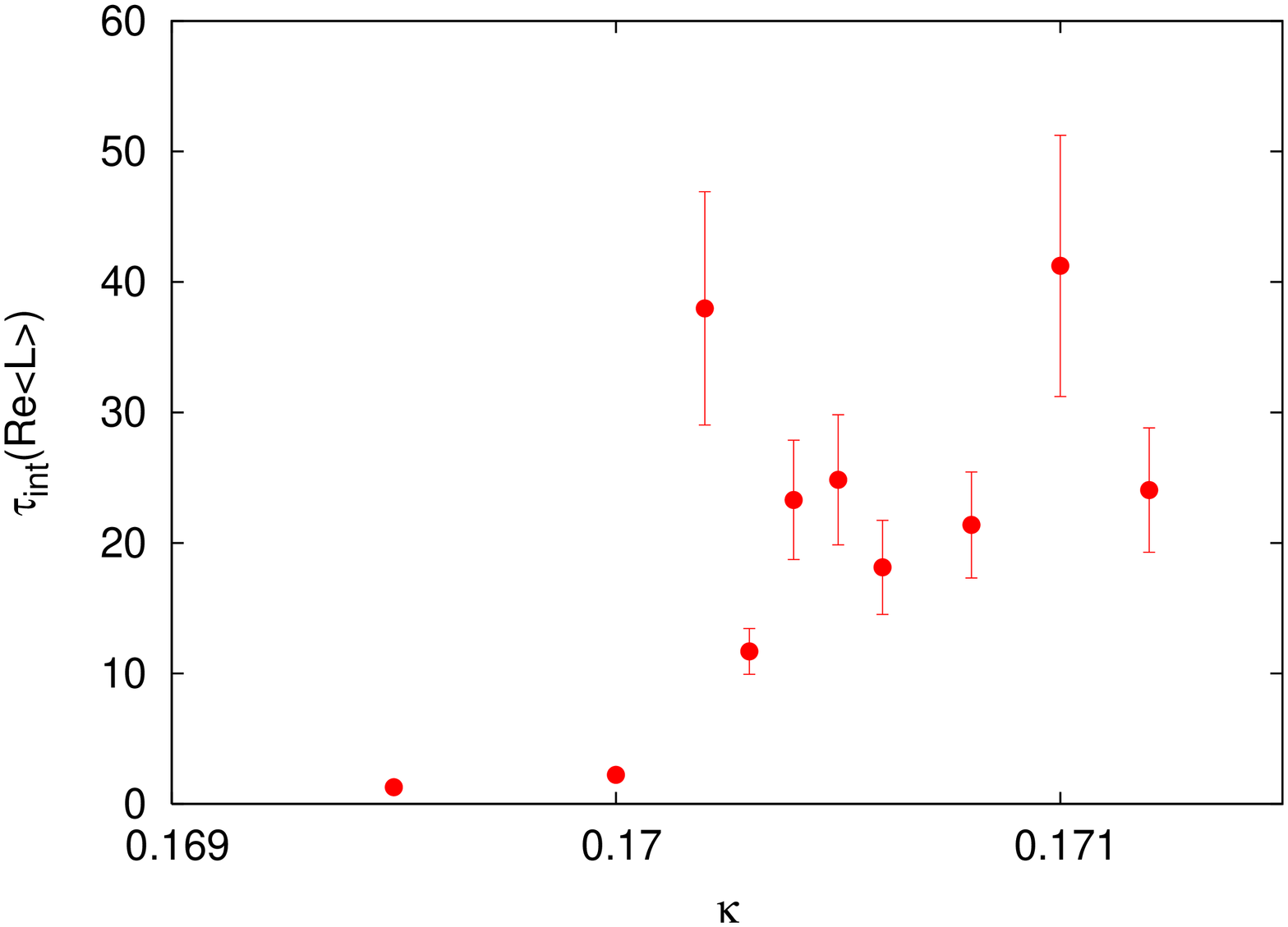}
        \includegraphics[width=0.35\textwidth,angle=-90]{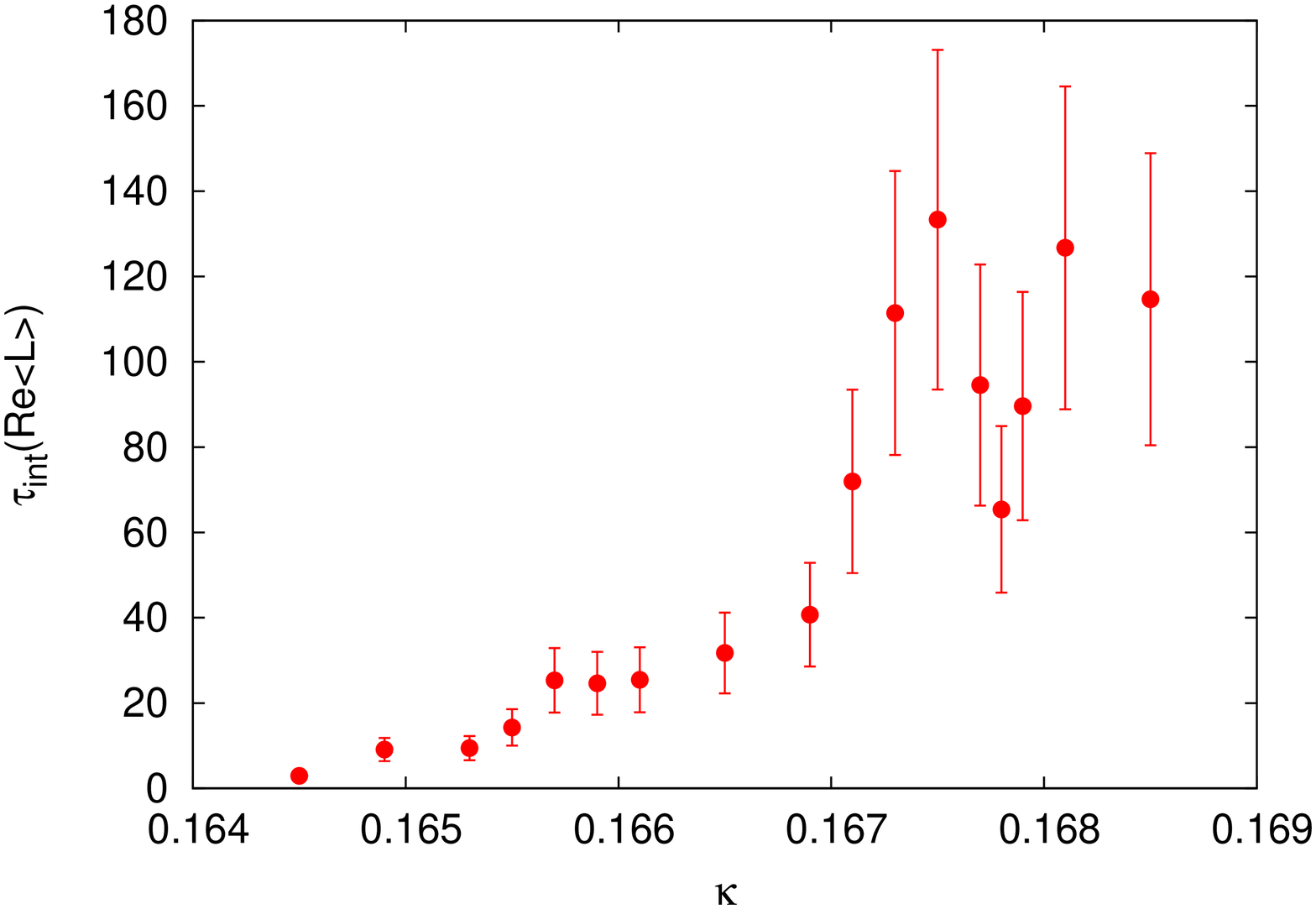}
    \caption[]{
      Integrated autocorrelation time for $\ReL$ as a function of $\kappa$ for
      $\beta=3.65,\,\munull=0.0068$ (left) and $\beta=3.75,\,\munull=0.0070$ (right).
      \label{fig:TauInt}}
\end{figure}

In this section we discuss our observations of the finite temperature transition, 
specifically its evolution along the $\beta-$axis. 
Despite the reasonable statistics of
typically $\mathcal{O}(15\text{K})$ trajectories
per data point, our results are still noisy and remain somewhat qualitative. 
This is also due to the fact that we are, most likely, dealing with a very soft
crossover rather than a true phase transition.

Let us start by reconsidering the right panel of Fig.~\ref{fig:MetaStab}.
The real part of the Polyakov loop rises significantly when 
$\kappa$ is increased, indicating the transition to a deconfined regime.
Upon crossing $\kcr$, however, it enters the metastability region discussed in 
the previous section and then drops again.
We interpret the rise and fall of the Polyakov loop along the qualitative expectations from
Fig.~\ref{fig:CCC}
as an indication for two finite temperature transitions: first a
confinement $\rightarrow$ deconfinement transition when approaching \kcr{}
from below, followed by a deconfinement $\rightarrow$ confinement 
transition at $\kappa>\kcr$. 
This finite temperature transition
is masked by the unphysical bulk transition around $\kcr{}$ if $\munull$ is
chosen too small. However, according to Sec.~\ref{subsec:WtmFermionPhaseStructure} 
this should be disentangled at larger values of $\beta$.

Indeed, choosing $\beta=3.65$ while keeping $\munull$ constant, we observe the behaviour
displayed in Fig.~\ref{fig:FiniteTb365}. 
The metastabilities are now suppressed and ${\langle\rm Re}(L)\rangle$
peaks before gently declining again.
Similar signals are obtained from other observables, with a sharper peak
in the pion norm marking the chiral transition at $\kappa_c$, a remnant from the 
zero temperature bulk transition. 

An inspection of the data points and their statistical uncertainties
reveals an asymmetry between $\kappa<\kcr$ and $\kappa>\kcr$. 
This is because for $\kappa>\kcr$ the spectrum of the twisted 
mass Dirac operator is shifted in the 
negative real direction, which implies the occurence of very small eigenvalues and
negative quark masses \cite{Jansen:2003ir}. 
The results of an autocorrelation analysis are shown in Fig.~\ref{fig:TauInt}.
The integrated autocorrelation time $\tauint$ of the Polyakov loop rises 
notably when passing $\kcr$, and it continues on a heightened plateau 
at negative quark masses.
Hence, while we believe to have good evidence for its existence,
the upper half cone of thermal transitions $\kappa_t > \kcr$ is not easily 
accessible by simulations.  

\begin{figure}[t]
  \includegraphics[width=0.35\textwidth, angle=270]{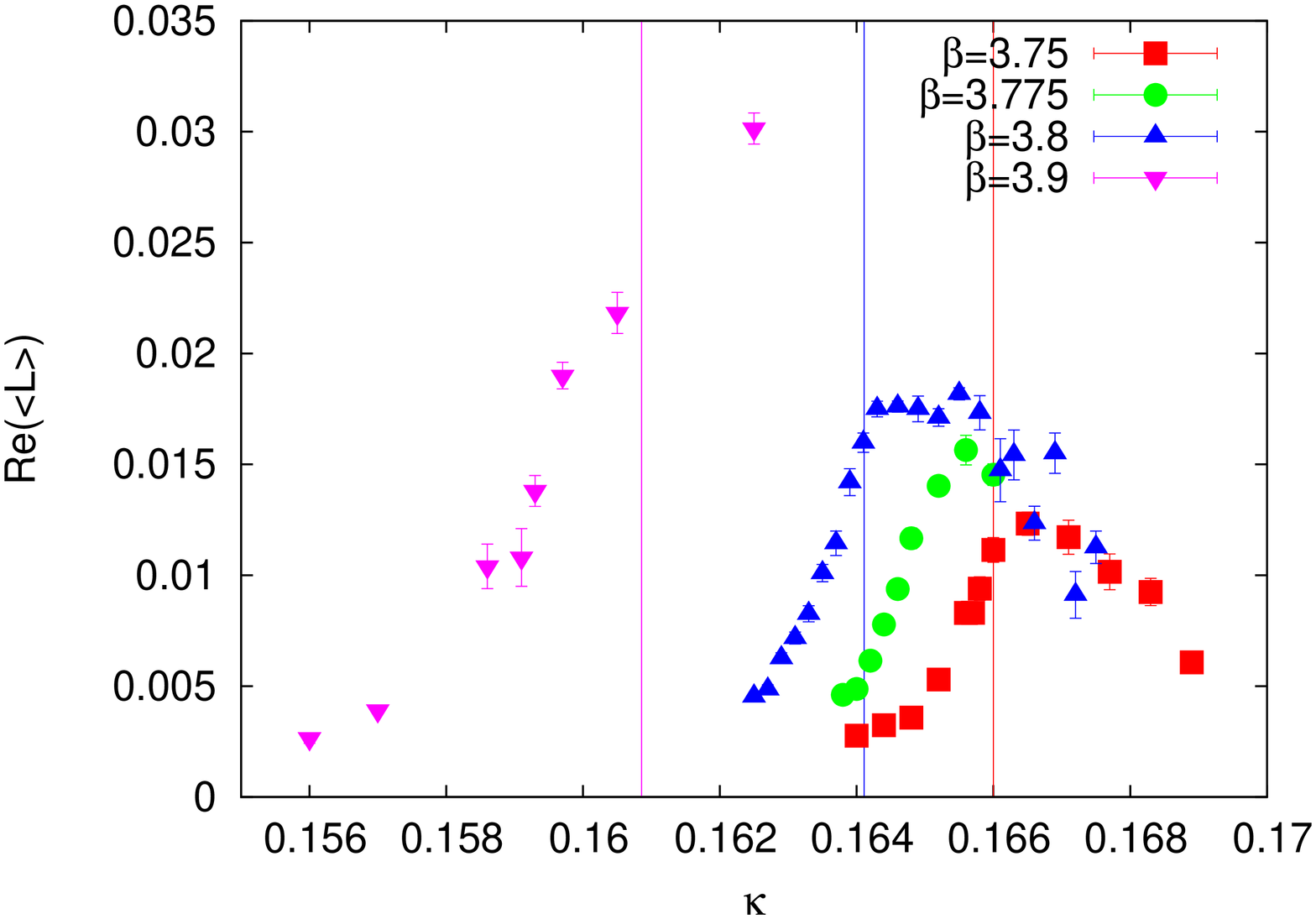}
  \includegraphics[width=0.35\textwidth, angle=270]{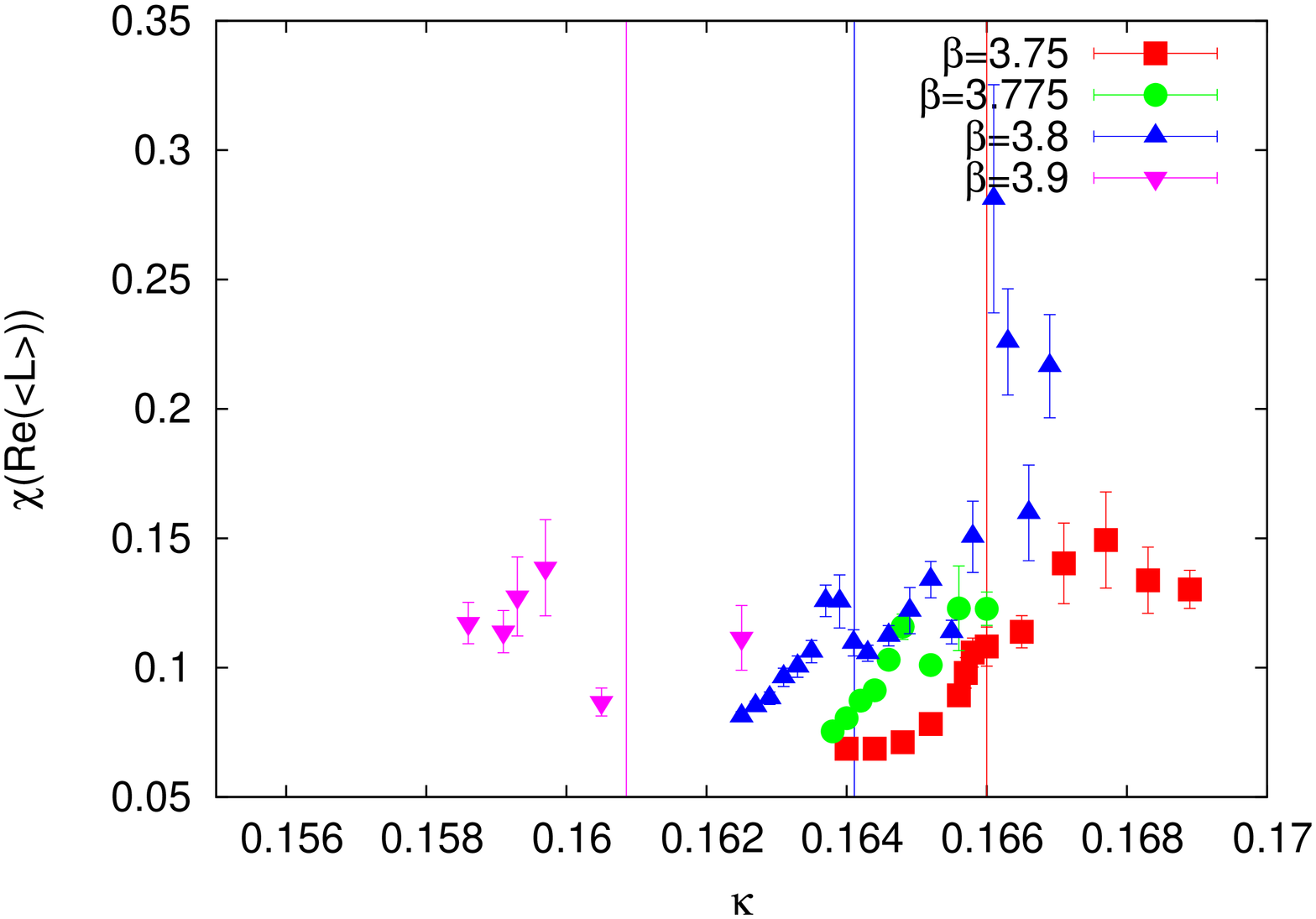}
  \caption[]{
    The real part of the Polyakov loop (left) and its susceptibility (right)
    for $\beta=3.75,\,3.775,\,3.8,\,3.9$ and $\munull=0.005$. 
    The vertical lines mark $\kcr(T=0,\beta)$ for $\beta=3.9,\,3.8,\,3.75$ from left to right.
    \label{fig:Ploop}}
\end{figure}
By a further increase of $\beta$ to $\beta\ge 3.75$ we 
approach the physically interesting weak bare coupling region.
Looking at the left panel of Fig.~\ref{fig:Ploop}, we once more find the real part of the 
Polyakov loop displaying a maximum around \kcr{}, which is consistent with the 
cone structure.
Note also that, in agreement with qualitative expectations, the cone is 
widening at larger $\beta$.
It is customary to locate phase transitions 
more precisely by generalised susceptibilities, which mark the points of maximal fluctuations.
These are shown on the right hand side of
Fig.~\ref{fig:Ploop} for the Polyakov loop.
The first transition appears to weaken with decreasing $\beta$ and turns into a rather flat 
shoulder at $\beta=3.75$.
Decreasing $\beta$ implies decreasing $T_c$, and since $T_c$ is an increasing function of
quark mass, the transitions for decreasing $\beta$-values belong to a sequence of decreasing
mass parameters.
Our signals are thus consistent with a weakening of the deconfinement transition 
when moving away from the
quenched regime.
The latter was roughly located for the tree-level Symanzik improved gauge action
by performing a series of $\beta-$scans at very large quark mass and $\munull=0.005$.
As expected \cite{Aoki:1995yf} the finite temperature transition line bends down
from the chiral chritical line towards the $\beta-$axis and hits the latter at a finite
value $\bqu\gtrsim 4.50(5)$ (cf.~Table \ref{tab:TransitionPoints}).

\begin{figure}[t]
      \includegraphics[width=0.35\textwidth, angle=270]{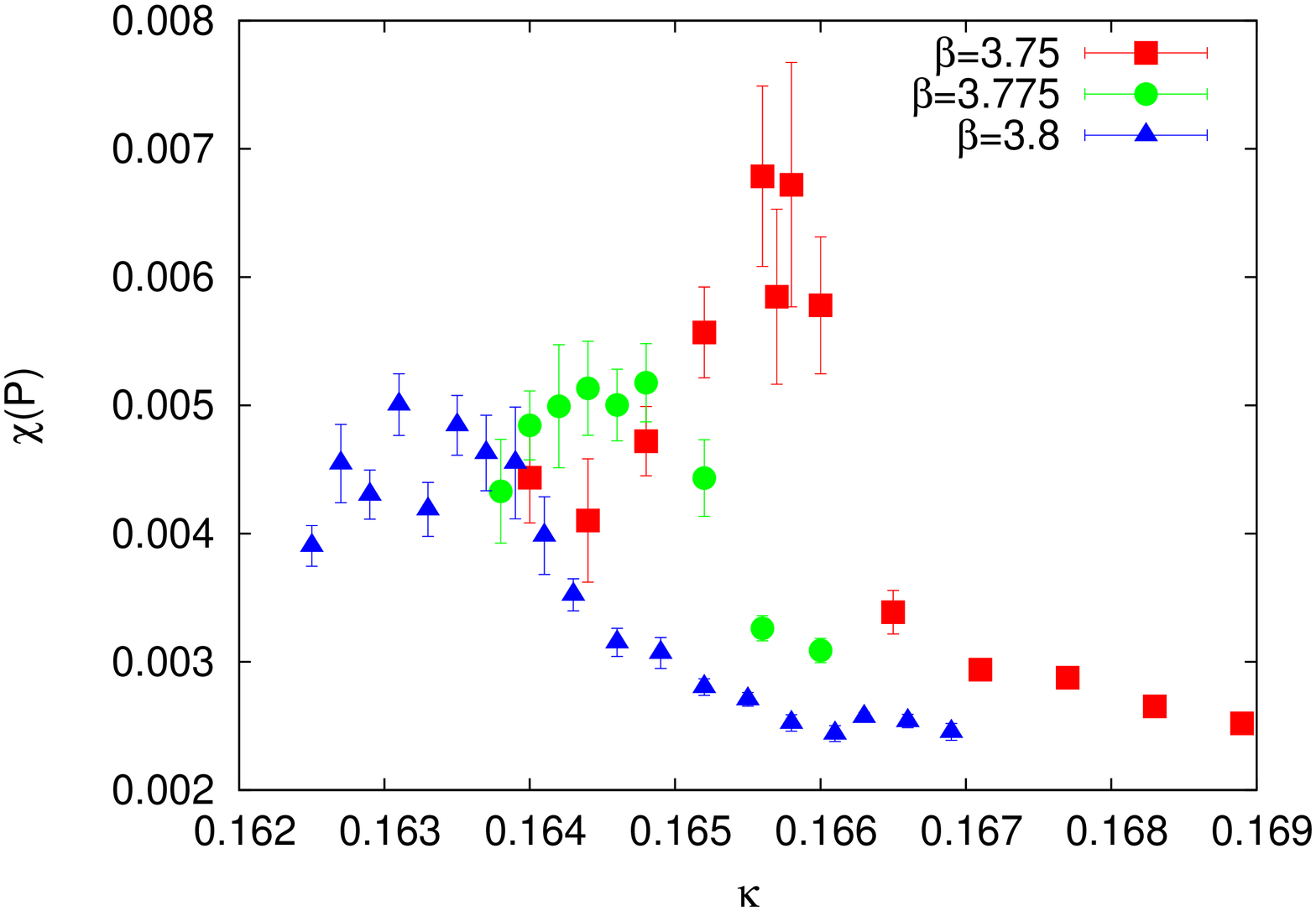}
      \includegraphics[width=0.35\textwidth, angle=270]{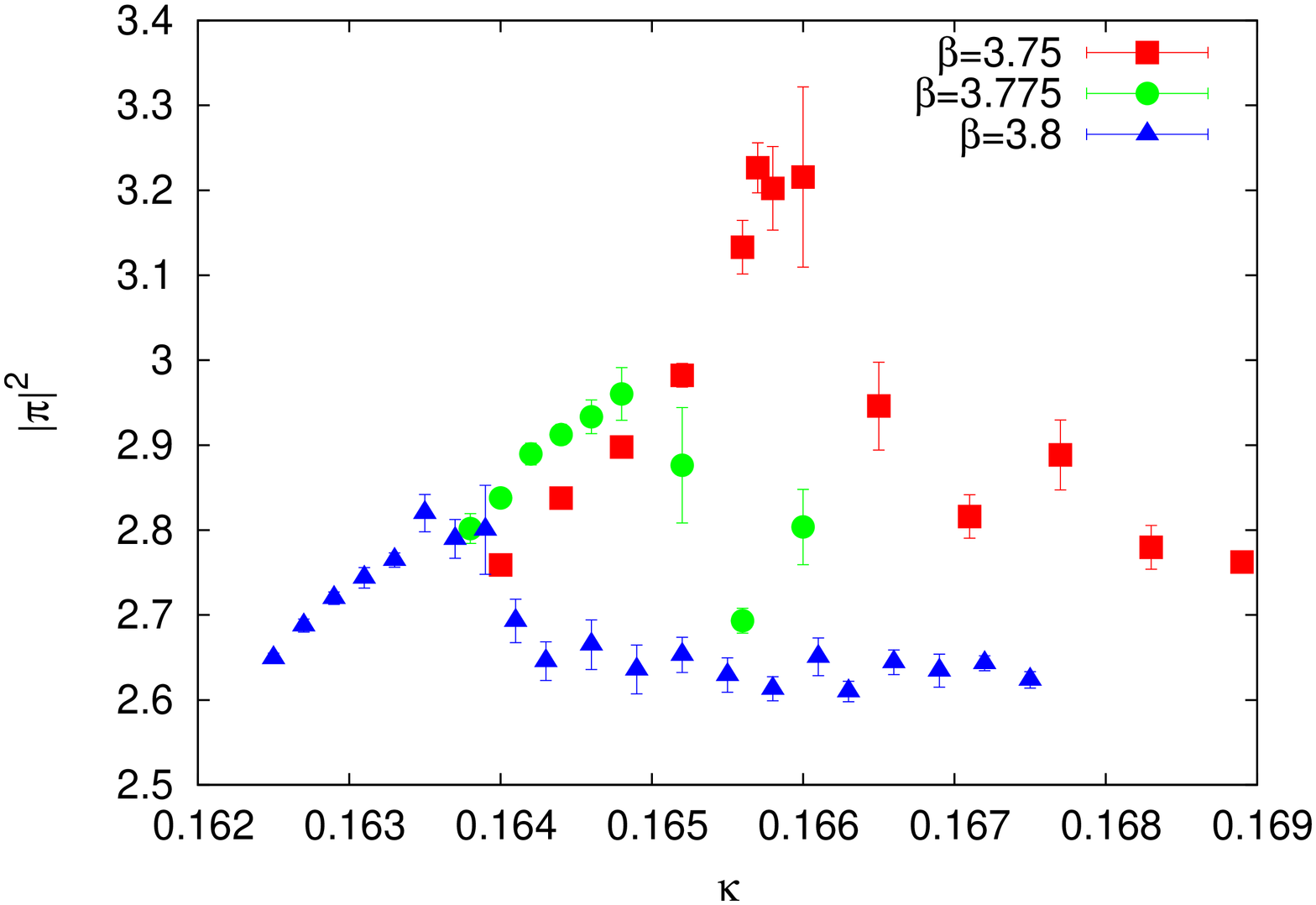}
    \caption[]{
      Plaquette susceptibility (left) and pionnorm (right) 
      for $\beta=3.75,\,3.775,\,3.8$ and $\munull=0.005$.
      \label{fig:PlaqsusPnorm}}
\end{figure}
\begin{figure}[t]
  \includegraphics[width=0.55\textwidth,angle=270]{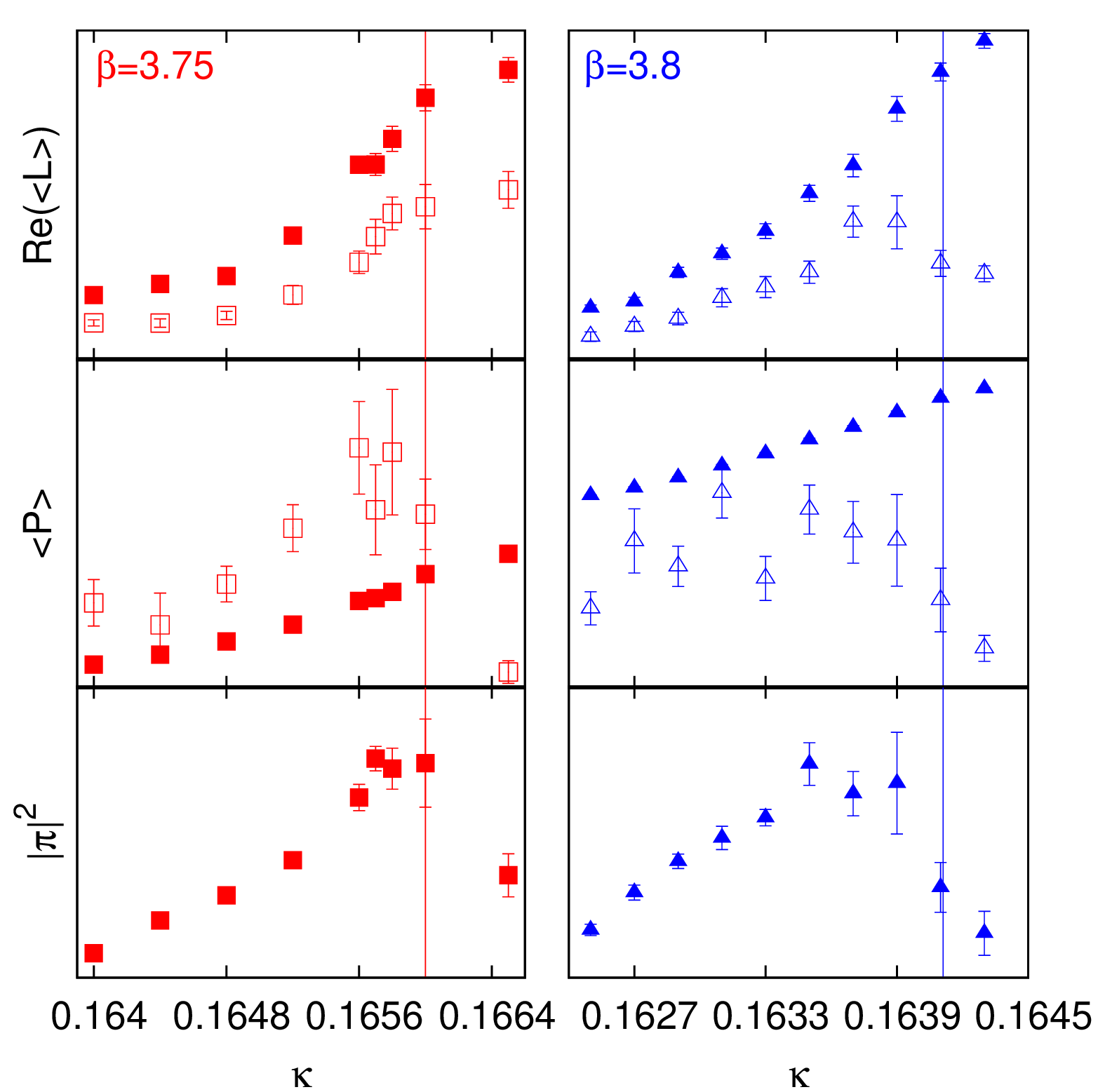}
  \includegraphics[width=0.35\textwidth,angle=-90]{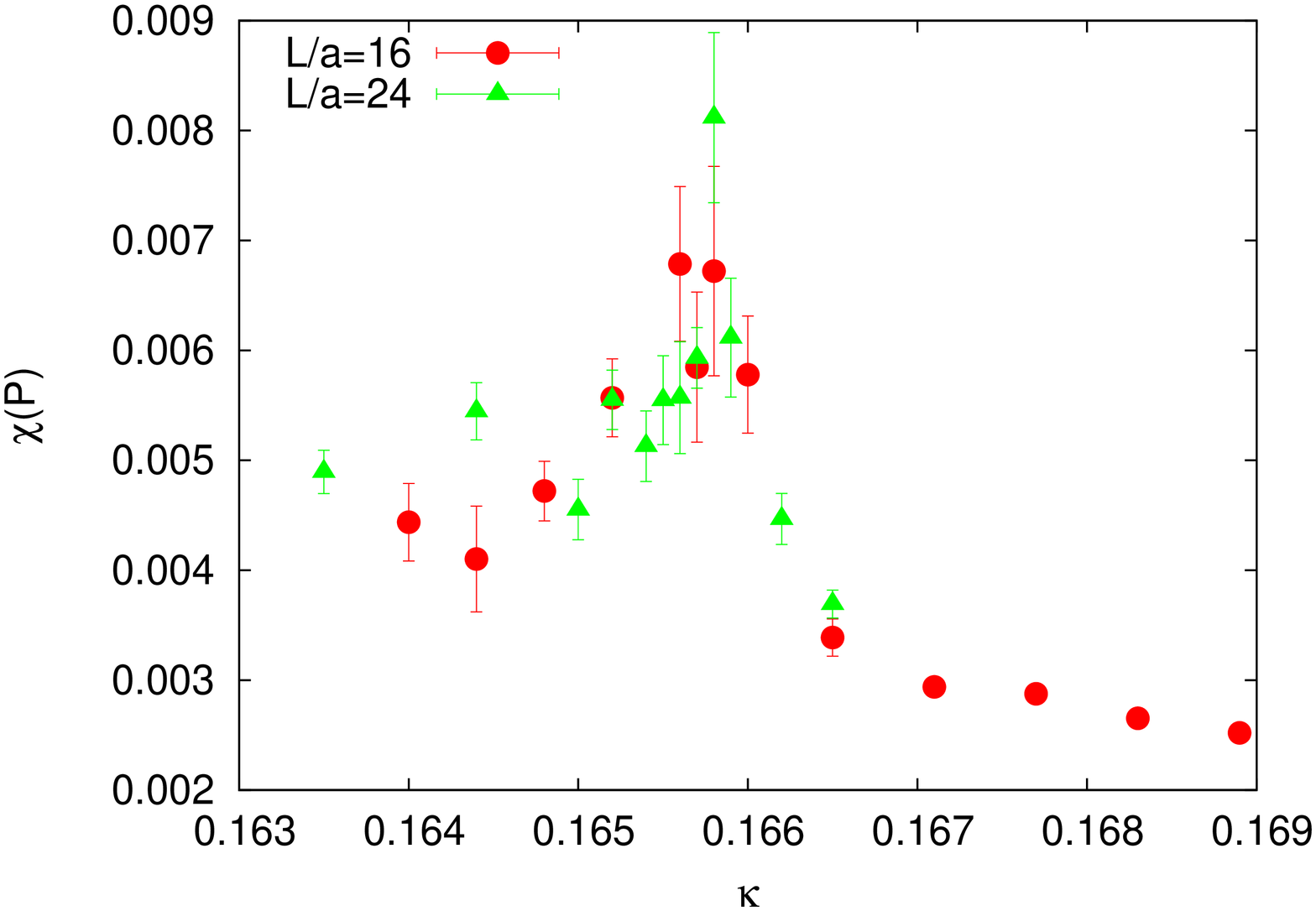}
\begin{minipage}{8cm}
  \caption{Real part of Polyakov loop (top panel), plaquette (middle panel)
    and pion norm (bottom)
    near $\kcr$ (marked by vertical lines). 
    For the first two observables expectation value (filled points) 
    and susceptibility (open) are shown. 
    \label{fig:kckt}}
\end{minipage}\hspace*{1cm}
\begin{minipage}{7cm}\vspace*{-7.5cm}
    \caption[]{
      Finite size scaling test for the plaquette susceptibility
      at $\beta=3.75$ and $\munull=0.005$.
      \label{fig:fss}}
\end{minipage}
\end{figure}

Fig.~\ref{fig:PlaqsusPnorm} shows that the qualitative picture obtained from 
the Polyakov loop is repeated in
the plaquette susceptibility (left) and the pion norm (right).
Note that in all cases the location of the first thermal transition is
at $\kappa_t<\kcr$, consistent with the qualitative expectations in 
Figs.~\ref{fig:aoki},~\ref{fig:CCC}. 
This becomes more obvious in Fig.~\ref{fig:kckt}, where we compare
in detail the relative position of $\kcr$ and $\kappa_t$ for
the confinement-deconfinement transition for $\beta=3.75,\,3.8$, as observed
in all three observables we consider. 
Whenever a peak in a susceptibility can be isolated, the value for $\kth$ can
be consistently determined from any observable by fitting Gaussian curves 
through the peaks.
Our estimated values
for the critical couplings are collected in Table \ref{tab:TransitionPoints}.

An additional run on a larger lattice 
does not  seem to result in a growth
of the susceptibility peaks, as displayed in Fig.~\ref{fig:fss},
a behaviour consistent with a crossover. Our more detailed study
at $\beta=3.75$, which will be reported in Sec.~\ref{subsec:kappamuplane},
will indeed indicate a pion mass over 400\,MeV, for which only a smooth crossover
behaviour is expected. 

The situation changes drastically for $\kappa>\kcr$: the persistence of the Polyakov
loop susceptibility at a high level --- a phenomenon we again relate to the 
peculiar dynamics and its algorithmic problems in this region of quark masses 
--- leaves the localisation of the second transition to be a delicate task.
However, we consider the descent of the Polyakov loop as a clear signal for
the deconfinement-confinement transition at $\kappa > \kcr$.
We further tested this interpretation by a $\beta-$scan with $\kappa=0.17$
above $\kappa_c$. 
%which should expose a thermal transition 
%if the upper branch indeed is to merge with the lower branch
%in the vicinity of $\kcr$. 
A signal for a transition
was indeed detected and is also given in Table \ref{tab:TransitionPoints}.

In contrast to the situation at $\kappa < \kcr$, these signals are
accompanied by a significantly weaker or altogether missing corresponding signal 
in other observables.
This demonstrates the influence of cut-off effects in the simulation of 
the lattice theory: while the thermal transition in the continuum is manifestly
invariant under changes of the twist angle, we find this invariance broken by 
lattice artefacts such that positive and negative values for the twist angle
can result in the two rather distinct signal scenarios we encountered.
Let us recall, however, that the physically interesting region is
the lower half of the cone with $\kappa\le\kcr$, 
where the signal for the transition can be located
and is consistent among different observables.

\subsection{Second branch of finite temperature transitions at large $\kappa$}
\label{subsec:FiniteTatLargeK}

\begin{figure}[t]
      \includegraphics[width=0.35\textwidth, angle=270]{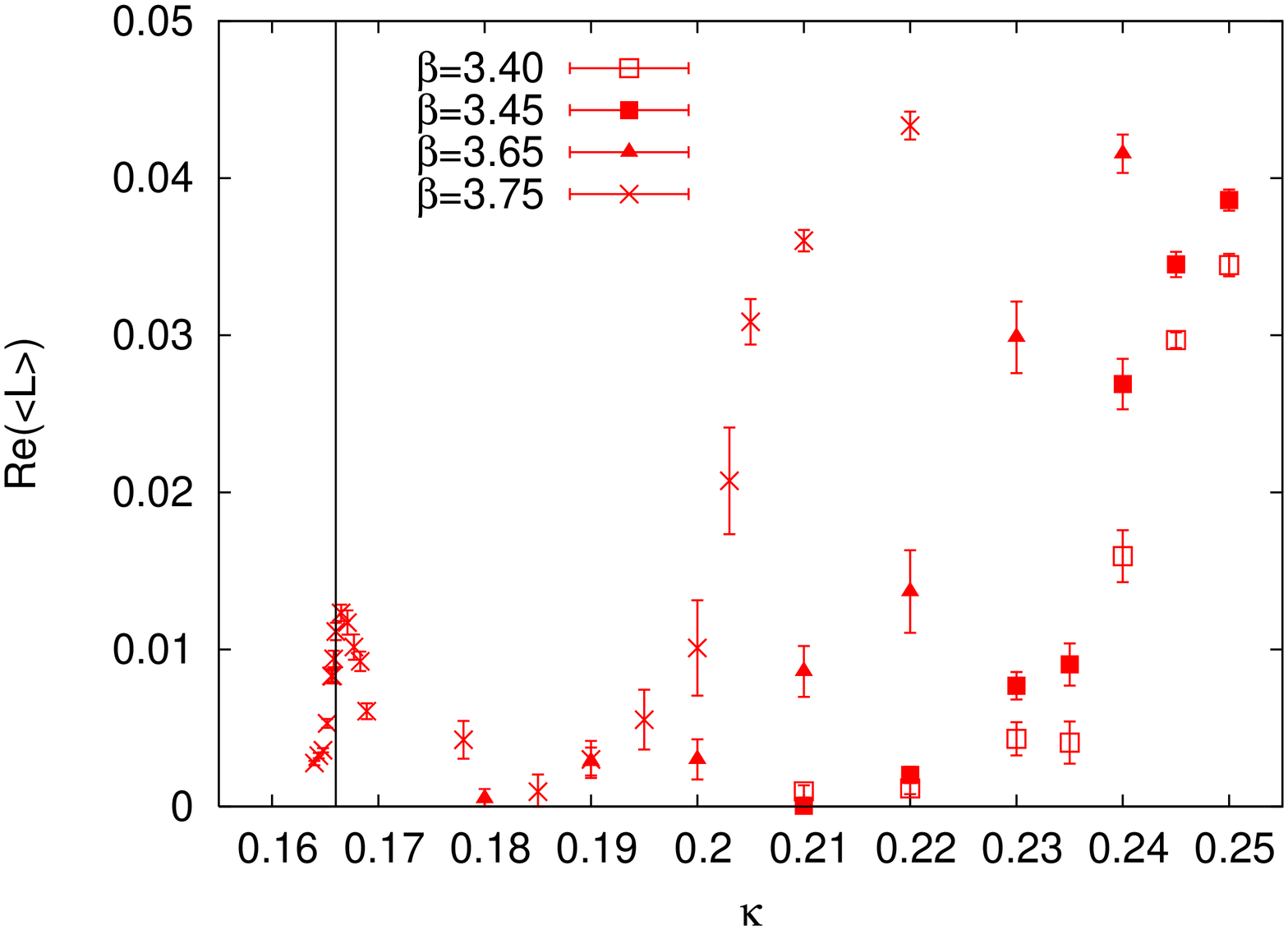}
      \includegraphics[width=0.35\textwidth, angle=270]{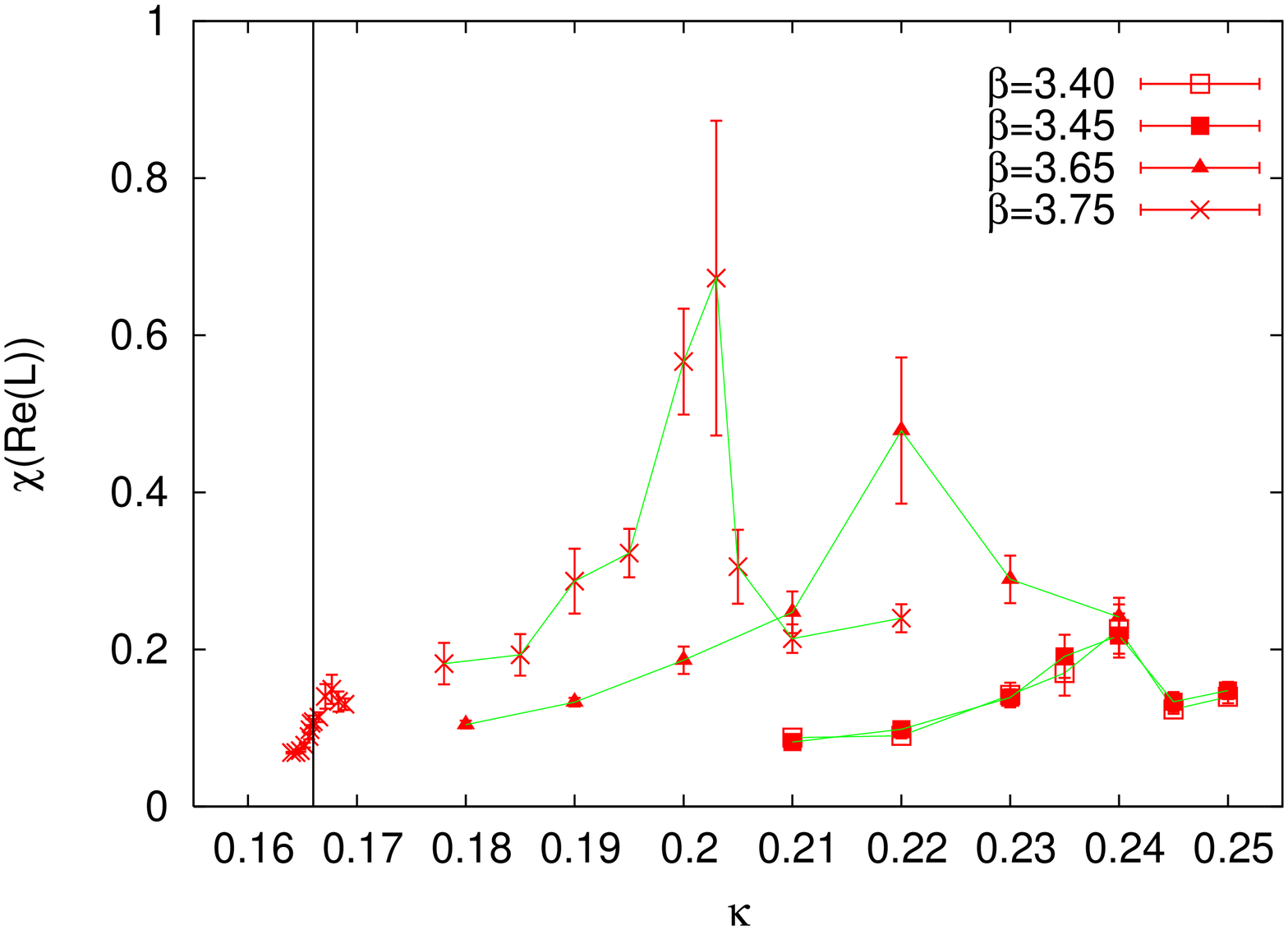}
    \caption[]{Real part of the Polyakov loop (left) 
and its susceptibility (right)
      for $\beta=3.4,\,3.45,\,3.65$ with $\munull=0.0068$ and $\beta=3.75$ 
      with $\munull=0.005$. The vertical line marks $\kcr(\beta=3.75)$. 
Lines on the right 
      are added for visual guidance.
      \label{fig:PloopLargeK}}
\end{figure}

\begin{figure}[t]
\begin{center}
\includegraphics[width=0.35\textwidth,angle=-90]{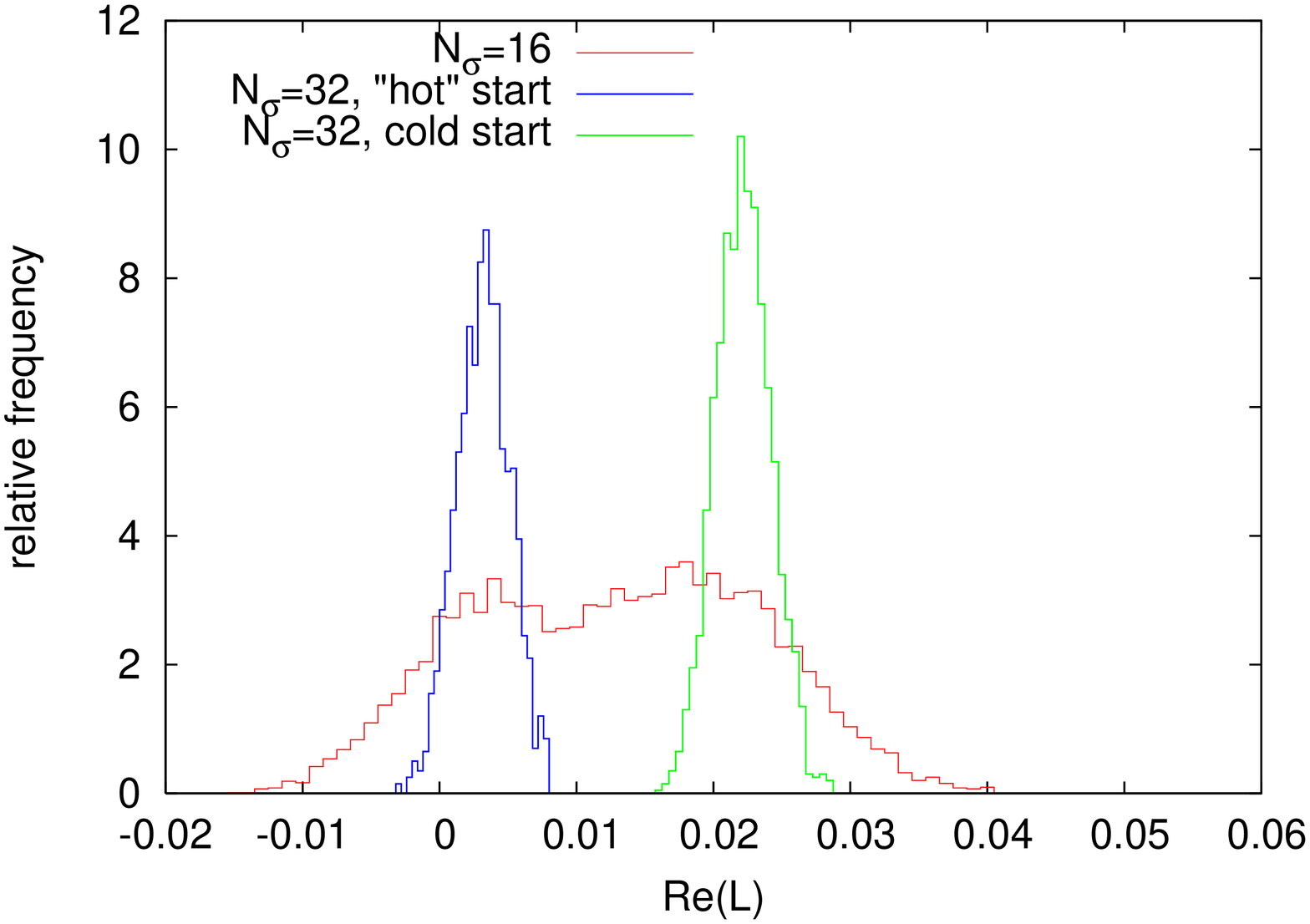}
\end{center}
    \caption[]{Distributions of \ReL{} for $\beta=3.6, \munull=0, \kappa=0.22$.
      \label{fig:PloopHist}}
\end{figure}

In order to gain a global perspective of the phase structure, 
we extended our investigation
 towards $\kappa \gg \kcr$, even though this region is not relevant for 
continuum physics. The
results for the Polyakov loop and its susceptibility of several 
$\kappa-$scans are collected
in Fig.~\ref{fig:PloopLargeK}. 
There appears to be another line of potential confinement $\rightarrow$ deconfinement 
transitions moving towards the chiral critical line with increasing $\beta$. 
The existence of such a branch of
transitions is consistent with the conjecture in Fig.~\ref{fig:CCC}: it matches
the line bending upwards from the upper 
end of the cone at large $\beta$ towards the first doubler
region. In order to give a rough idea of the distance between the two branches 
we present the data over a large $\kappa$-range including also the data near $\kcr$ for 
$\beta=3.75$.
The distributions of the  Polyakov loop in Fig.~\ref{fig:PloopHist}
and especially the pronounced metastability found when simulating on a 
spatially enlarged 
lattice with hot and cold starts 
suggest a possible first order nature of the transition 
in this parameter regime.

\subsection{Thermal transition line at fixed $\beta$}
\label{subsec:kappamuplane}

\begin{figure}[t]
\includegraphics[width=0.35\textwidth,angle=-90]{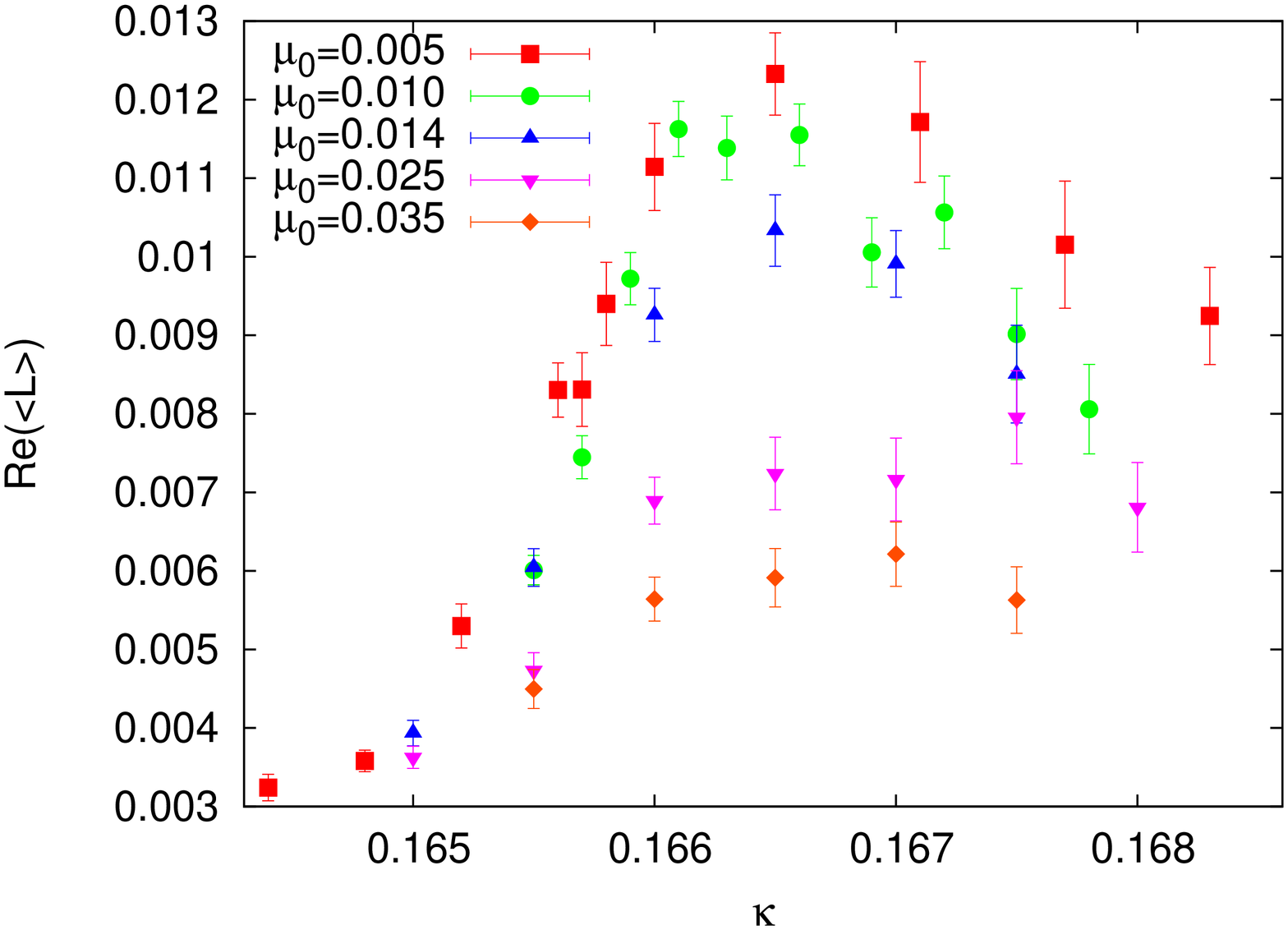}
\includegraphics[width=0.35\textwidth,angle=-90]{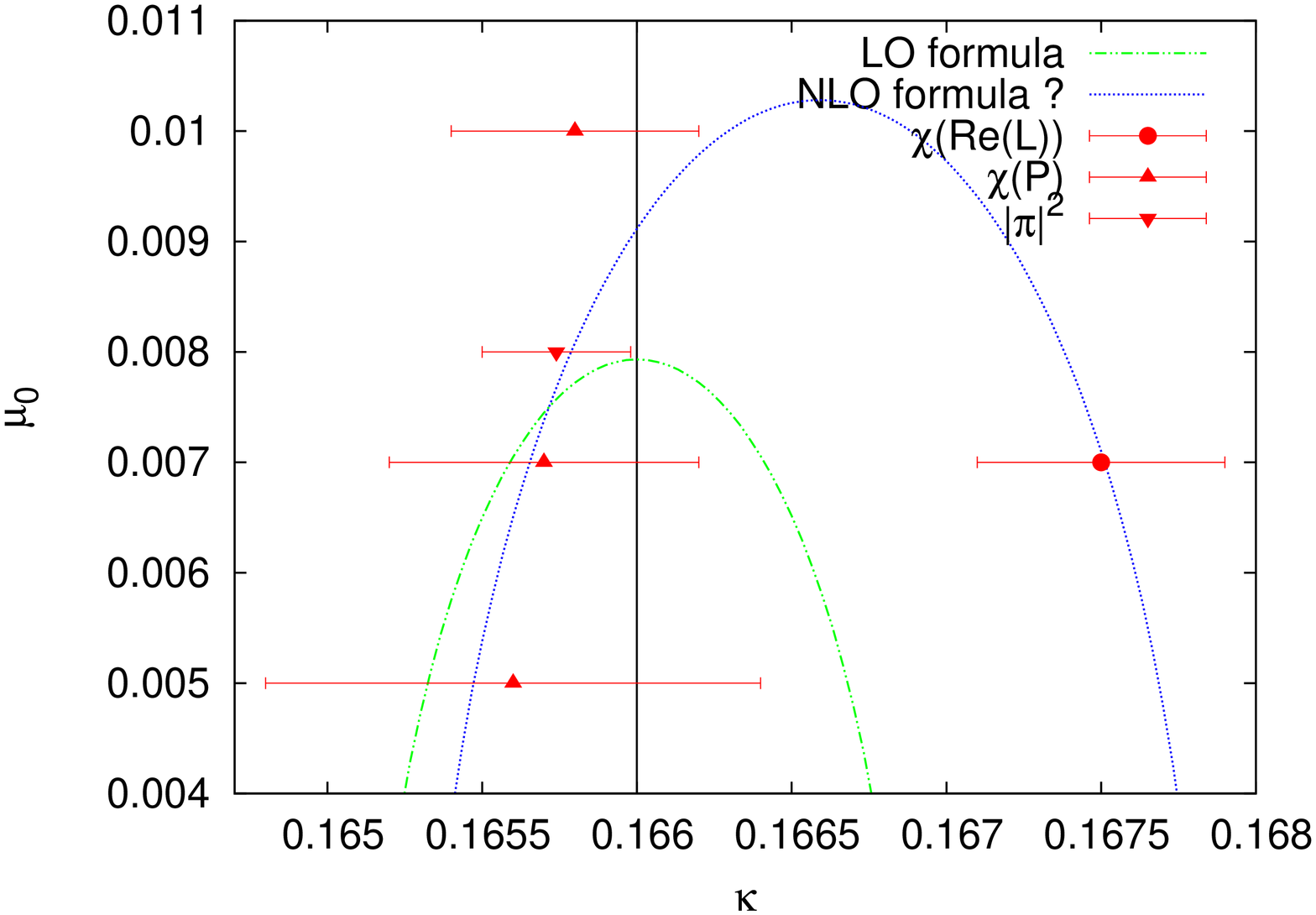}
  \caption[]{
      Left: Polyakov loop expectation value for different values of $\mu_0$. 
      For $\mu_0\geq 0.025$, the transitions are lost. 
      Right: Data points represent the phase boundary at $\beta=3.75$, 
      estimated from the peaks of susceptibilities.
      The leading order part of Eq.~(\ref{eq:mpi}) does not fit the data, 
      the NLO-formula contains
      undetermined constants that have been estimated by order of magnitude only.
      The vertical line marks $\kappa_c(\beta=3.75)$, i.e.~maximal twist.
      \label{ellipse}}
\end{figure}

Having established the overall qualitative phase structure, 
let us now focus on the shape of the thermal
transition surface in a plane of fixed $\beta$.
To this end we supplement our $\kappa$-scan at $\beta=3.75$, $\mu_0=0.005$ from
Figs.~\ref{fig:Ploop}, 
\ref{fig:PlaqsusPnorm} by additional ones at larger 
$\mu_0$ values.
The qualitative behaviour remains unchanged, 
but the two thermal transitions should approach each
other with increasing $\mu_0$, as expected for a slice of a conical shape. 
Eventually, for large values of $\mu_0$ one is outside the cone 
and the transition signal should be lost. This is shown in Fig.~\ref{ellipse} (left),
where the two transitions can be seen in the rise and fall of the real part of the 
Polyakov loop. For the largest twisted masses $\mu_0=0.025,0.035$ this behaviour 
is definitely lost.

\begin{figure}[t]
\centering
\includegraphics[width=0.34\textwidth,angle=-90]{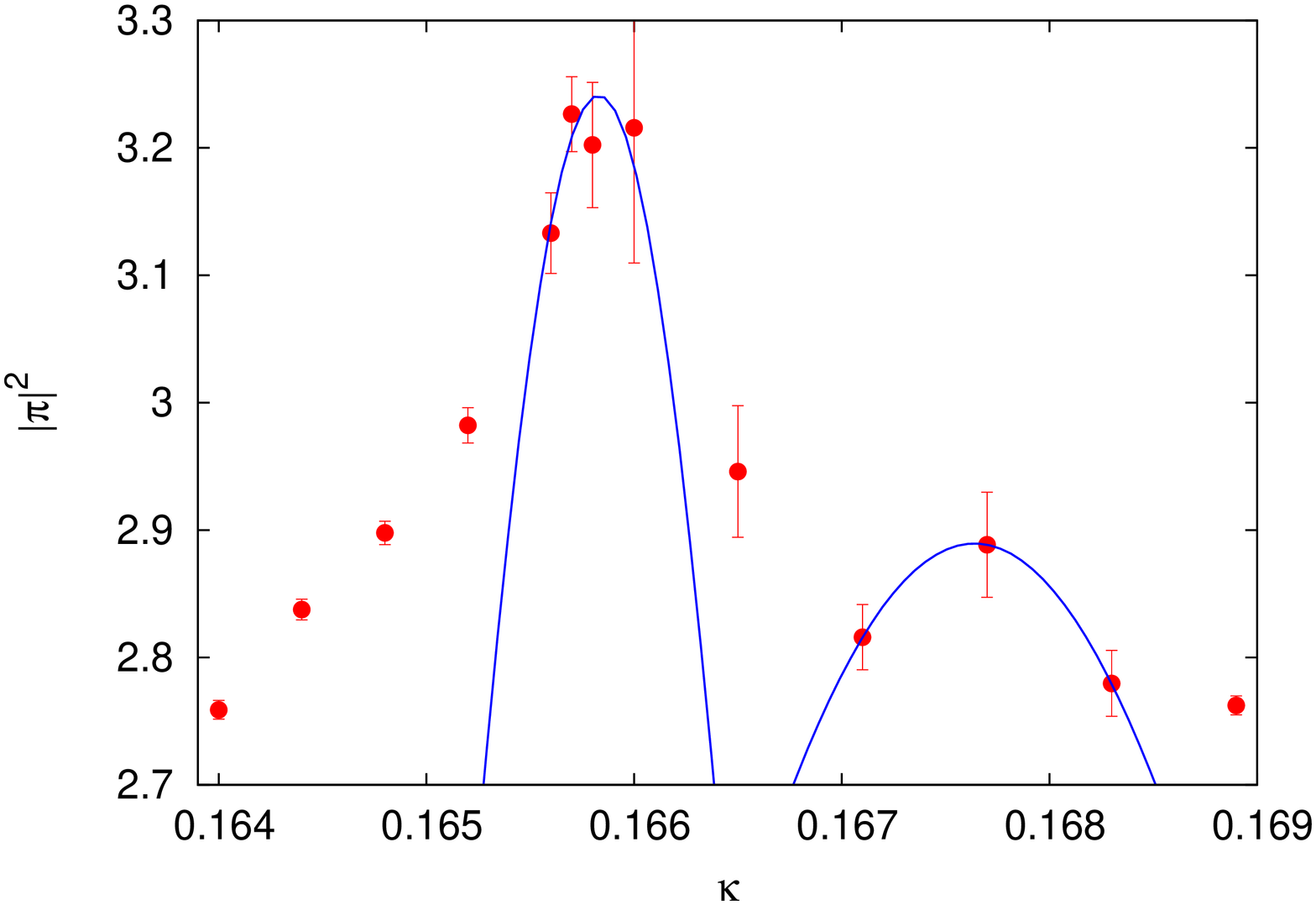}\hfill
\includegraphics[width=0.34\textwidth,angle=-90]{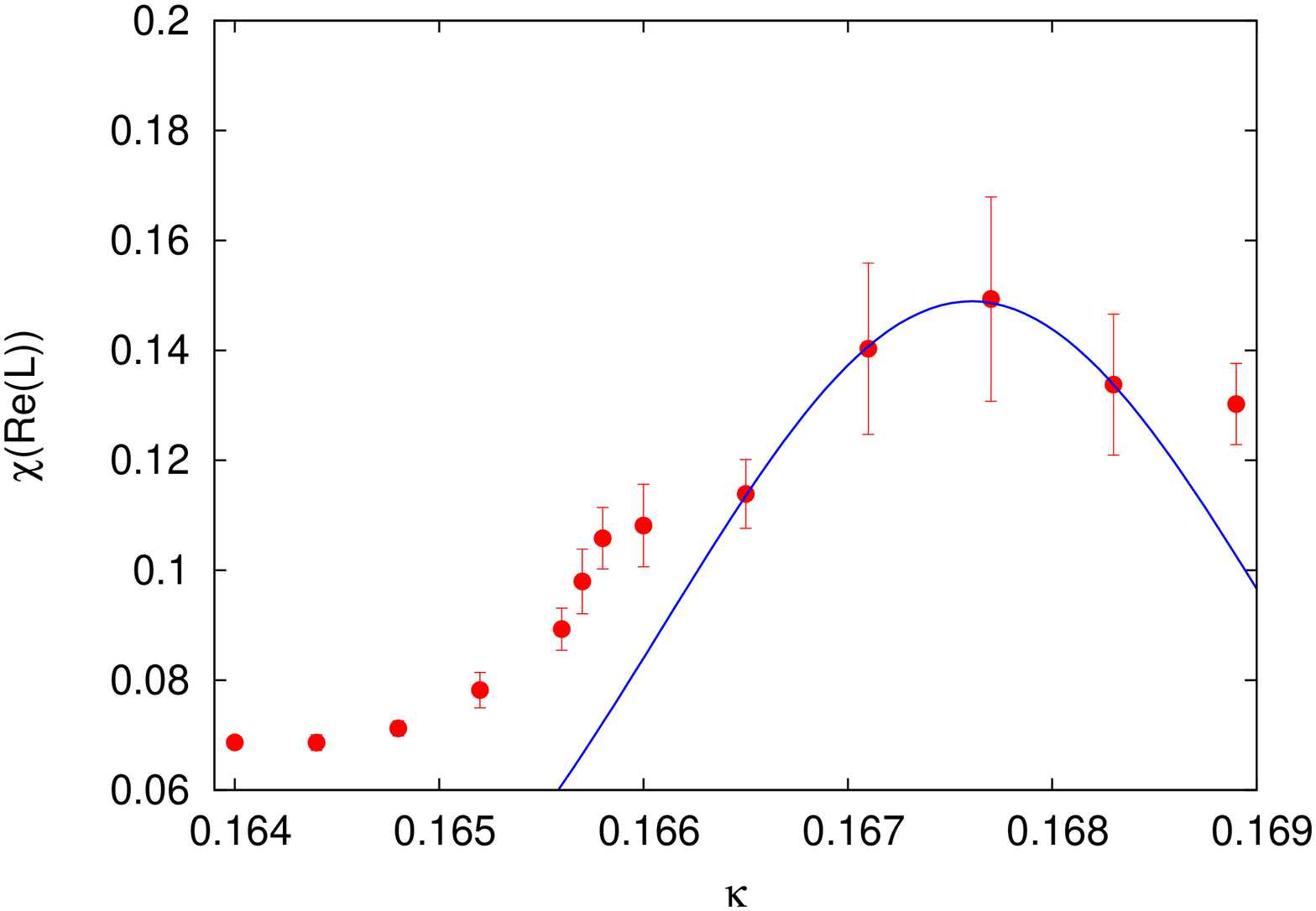}
  \caption[]{
      Pion norm (left) and Polyakov loop susceptibility (right) 
      showing two subsequent
      transitions at $\beta=3.75,\munull=0.005$.
      \label{peaks}}
\end{figure}

In order to locate the transition points more precisely,
we have attempted to fit Gaussian curves through
the peaks in the susceptibilities, as shown in the examples in Fig.~\ref{peaks}.
In particular, we may use the clear first transition in Fig.~\ref{peaks} (left)
for a rough estimate of the corresponding pion mass. According to
Table \ref{tab:TransitionPoints} the thermal transition point $\kappa_t$ is in this
case very close to maximal twist $\kappa_c(\beta=3.75)$, for which $\mu_0=0.005$ corresponds
to a pion mass $\sim 400$ MeV \cite{ETMC:private}. Since smaller $\kappa_t$ implies a slightly
larger mass, the thermal cone corresponds to a pion $\gtrsim 400$ MeV.
 
As was discussed before, different observables
display signals of different quality, and our data are not precise enough
to do an unambiguous determination of both transitions in all observables. 
Moreover, the underlying physical transition 
is a broad crossover. For parameter values where entry and exit of the deconfined
phase are close, the signals overlap and Gaussian fits do not work reliably,
cf.~Fig.~\ref{peaks} (right).
Nevertheless, in those cases that could be resolved, 
the peak positions are consistent between all 
observables, so we have used the observable with the strongest signal in each case.
This results in a qualitative picture of
the phase boundaries as shown in Fig.~\ref{ellipse} (right).


As discussed in Sec.~\ref{subsec:CCC}, in principle the cone of thermal transitions
corresponding to a given pion mass at fixed $\beta$ can be described by lattice
chiral perturbation theory, Eq.~(\ref{eq:mpi}), provided the mass and cut-off effects
are sufficiently small. We observe that the leading order formula with only the first term
in Eq.~(\ref{eq:mpi}) is unable to accomodate our data, even if an unknown constant
is left open as a fitting parameter. This is not surprising given the large pion
mass corresponding to the transition at $\beta=3.75$. 
For the NLO curve in Fig.~\ref{ellipse} (right)
we employed some very rough estimates for the unknown parameters in Eq.~(\ref{eq:mpi}) 
to check for consistency with that formula.
It would now be interesting to really determine
the unknown constants in the full NLO-formula Eq.~(\ref{eq:mpi}), in order to see
whether it correctly describes the data. 

\section{Discussion and conclusions}
\label{sec:Discussion}

\begin{figure}[p]
\centering
\includegraphics[width=0.55\textwidth,angle=-90]{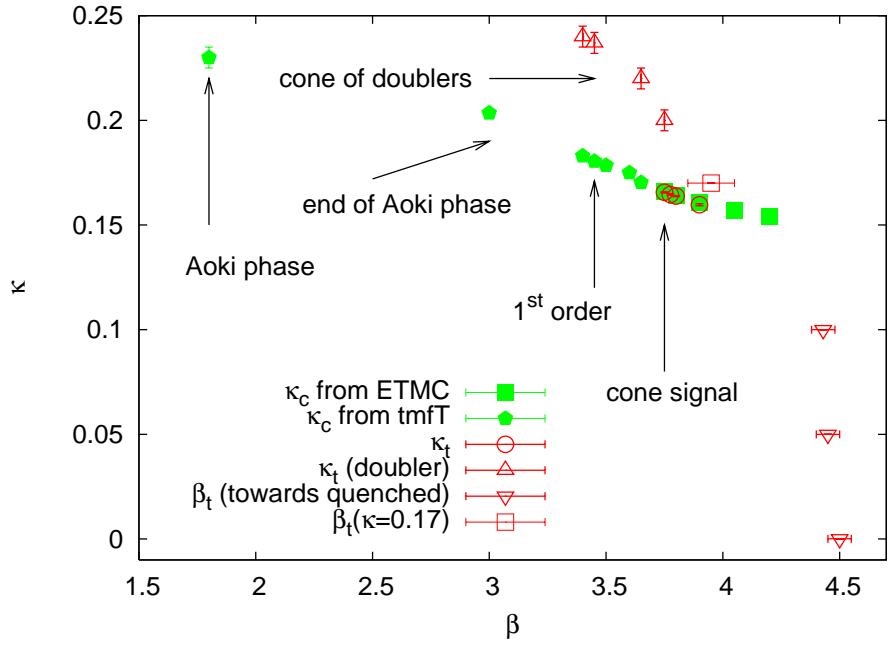}
\caption
    {Summary of numerically obtained transition points for $0\le \munull \lesssim 0.007$.
      The data for $\kcr(\beta)$, $\beta\ge 3.75$ are
      taken from \cite{Urbach:2007rt,ETMC:private}. 
      \label{fig:DataSummary}}
\end{figure}

We investigated the phase structure of two-flavour Wilson twisted mass fermions
at non-zero temperature with special interest in the thermal transition close to
the zero temperature chiral critical line.
Our estimates for transition points from numerical simulations are summarised in
Fig.~\ref{fig:DataSummary}, augmented by data for $\kcr(\beta\ge 3.75)$
from the ETM collaboration~\cite{Urbach:2007rt,ETMC:private} and the results of a 
rough tracking of the thermal line towards the quenched limit.

\begin{figure}[p]
\centering
\includegraphics[width=.8\textwidth]{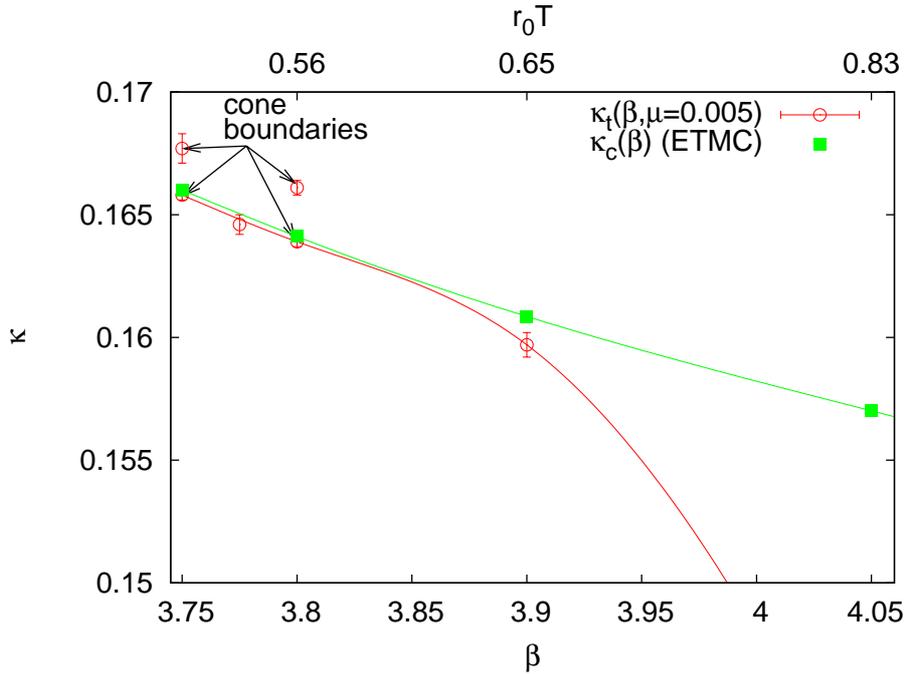}
    \caption[]
    {The region $\beta\ge3.75$. The lines are splines to guide the eye where the quenched and zero coupling limits have been taken into account.\label{fig:datasummaryzoom}}
\end{figure}

In Fig.~\ref{fig:datasummaryzoom}, we present an 
expanded view of the scaling region. The values of $r_0 T$ 
are shown along the upper horizontal axis (cf. reference \cite{Farchioni:2005ec} for the value
of $r_0/a$ in the chiral limit). Note that for improved staggered and Wilson fermions, 
extrapolations of the transition temperature to the chiral limit give 
$r_0 T_c\sim 0.45$ \cite{Bornyakov:2007zu,Cheng:2006qk},
much smaller than our explored range.
This is consistent with our large value of the pion mass and shows that larger $N_t\geq 10$
are required to approach the chiral limit of two flavor QCD.


Using the tree-level Symanzik improved Wilson gauge action, we produced substantial
results in support of both the existence of an Aoki phase at strong coupling and
the realisation of the Sharpe-Singleton scenario adjacent to the Aoki phase, with a bulk
transition surface narrowing
towards the weak coupling region. Our results for the thermal
transition are consistent with the conical structure proposed in
\cite{Creutz:2007fe}. 

From our extensive studies exploring a wide range of parameter space
we may then conclude that the global phase structure
of twisted mass lattice QCD is understood. Moreover, we have shown that for lattices 
with $N_\tau\geq 8$ this phase structure is beginning to show the features expected in the 
continuum. In particular, it is then feasible to tune to maximal twist by setting 
$\kappa=\kappa_c(\beta)$,
thus ensuring automatic $O(a)$-improvement,
and investigate the thermal transition in the $\{\beta,\mu_0\}$-plane without running into
unphysical phases. A recent perturbative 
study has shown that in this case maximally twisted Wilson fermions
show comparable scaling behaviour in the high temperature region as standard 
staggered fermions \cite{Philipsen:2008gq}, while such simulations should not be any 
more complicated than
those with untwisted Wilson fermions. It would thus be interesting to apply the twisted mass
formulation to systematic investigations of QCD thermodynamics. 

With results in this direction we hope to come up soon. Employing the twisted mass
formulation we would like to add an independent universality check to the
resolution of the ongoing controversary concerning the value of the critical
temperature(s) and the issue of a possible separation between the deconfining
transition and the chiral transition (for a recent paper on this subject see
\cite{Bornyakov:2009qh} and papers cited therein). We are confident that the automatic
$O(a)$ improvement will be technically advantageous for this task. Moreover, the extension
to the more realistic case $N_f=2+1+1$ seems to be straightforward (see \cite{Baron:2008xa}).
For studying the thermodynamic state equation, i.e. for computing the pressure
or the energy density, as well as the mesonic screening masses as a
function of the temperature we do not expect serious complications related to the
presence of the twisted mass term. However, in as far the explicit flavor symmetry
breaking will affect the thermodynamic behavior remains to be seen.

\vspace*{1cm}
\noindent
{\bf Acknowledgements:} We thank Mike Creutz, Steve Sharpe, Andr{\'e} Sternbeck and Carsten Urbach
for many useful discussions
and again Carsten Urbach for his supply, maintenance and support of the HMC
code. We also thank David Schulze for conducting the finite size check on $L=24$.
Most simulations have been done on the
apeNEXT computers in Rome and DESY-Zeuthen. This work has been supported in part by the DFG
Sonderforschungsbereich/Transregio SFB/TR9.
O.~P. and L.~Z. acknowledge support by the DFG project PH 158/3-1.
E.-M.~I. was supported by DFG under contract FOR 465 / Mu932/2
(Forschergruppe Gitter-Hadronen-Ph\"anomenologie).

\newpage
\clearpage

\appendix

\section*{Appendix: Simulation parameters}
%% \label{sec:appSimulationParameters}

    In Table~\ref{tab:SimulationRuns} we give a survey of all parameter 
    sets for the full
    scan of the phase diagram, ordered according to $\beta$. 
    The explored $\kappa$ range is given as well as the values of the
    twisted mass parameter $\mu_0$. 
    The last four entries show the $\beta$ range for scans performed
    at fixed $(\kappa,\mu_0)$. 

\begin{table}[h!]
  \begin{center}
    \begin{tabular}{ccc}
      $\beta$ & $\kappa$ & $\mu_0$  \\
      \hline
      \hline
       $1.8$ & $0.18\le\kappa\le 0.25$ & $0.010/(2\kappa)$ \\
       $1.8$ & $0.18\le\kappa\le 0.25$ & $0.005/(2\kappa)$ \\
       $1.8$ & $0.18\le\kappa\le 0.25$ & $0.003/(2\kappa)$ \\
      \hline
       $3.0$ & $0.18  \le \kappa \le 0.22$  & $0.010/(2\kappa)$ \\
       $3.0$ & $0.203 \le \kappa \le 0.204$ & $0.008/(2\kappa)$ \\
       $3.0$ & $0.203 \le \kappa \le 0.204$ & $0.005/(2\kappa)$ \\
       $3.0$ & $0.203 \le \kappa \le 0.204$ & $0.003/(2\kappa)$ \\
       $3.0$ & $0.2035$ & $0.0019/(2\kappa)$\\
      \hline
       $3.4$ & $0.16 \le \kappa \le 0.24$ & $0.010/(2\kappa)$ \\
       $3.4$ & $0.16 \le \kappa \le 0.20$ & $0.005/(2\kappa)$ \\
       $3.4$ & $0.16 \le \kappa \le 0.1875$ & $0.0025/(2\kappa)$ \\
      \hline
       $3.45$ & $0.179 \le \kappa \le 0.1835$ & $0.0025/(2\kappa)$ \\
      \hline
       $3.65$ & $0.169 \le \kappa \le 0.176$ & $0.0068$ \\
      \hline
      $3.75$  & $0.1640 \le \kappa \le 0.1689$ & $0.005$ \\
      $3.775$ & $0.1640 \le \kappa \le 0.1660$ & $0.005$ \\
      $3.8$   & $0.1627 \le \kappa \le 0.1675$ & $0.005$ \\
      $3.9$   & $0.1550 \le \kappa \le 0.1625$ & $0.005$ \\
      \hline
      $3.75$  & $0.1645 \le \kappa \le 0.1685$ & $0.007$ \\
      $3.75$  & $0.1649 \le \kappa \le 0.1677$ & $0.008$ \\
      $3.75$  & $0.1655 \le \kappa \le 0.1678$ & $0.010$ \\
      $3.75$  & $0.1655 \le \kappa \le 0.1680$ & $0.012$ \\
      $3.75$  & $0.1655 \le \kappa \le 0.1670$ & $0.014$ \\
      $3.75$  & $0.1650 \le \kappa \le 0.1680$ & $0.025$ \\
      $3.75$  & $0.1655 \le \kappa \le 0.1675$ & $0.035$ \\
      \hline
      $3.4$ & $0.20 \le \kappa \le 0.26$ & $0$ \\
      $3.5$ & $0.20 \le \kappa \le 0.26$ & $0$ \\
      \hline
      $3.4$ & $0.21 \le \kappa \le 0.25$ & $0.0068$ \\
      $3.45$ & $0.21 \le \kappa \le 0.25$ & $0.0068$ \\
      $3.65$ & $0.18 \le \kappa \le 0.24$ & $0.0068$ \\
      $3.75$ & $0.178 \le \kappa \le 0.22$ & $0.005$ \\
      \hline
      $3.75 \le \beta \le 4.2$ & $0.17$ & $0.005$ \\
      \hline
      $3.7 \le \beta \le 5.0$ & $0.0001$ & $0.005$ \\
      $3.9 \le \beta \le 4.6$ & $0.05$ & $0.005$ \\
      $4.0 \le \beta \le 4.5$ & $0.1$ & $0.005$ \\
      %M1  & $3.75$  & $\kcr$ & scanned \\
      %M2  & $3.8$   & $\kcr$ & scanned \\
      \hline
      \hline
      %\hline
    \end{tabular}
    \caption{Parameter sets for the simulations.
    \label{tab:SimulationRuns}}
  \end{center}
\end{table}

    In Table~\ref{tab:TransitionPoints} we list the locations 
    of the identified thermal transitions
    $\kappa_t$ and the observables, from which they were extracted. For the
    large $\beta$ values the estimate is based on the distributions of real
    and imaginary parts of the Polyakov loop. If two values of $\kappa_t$ are
    given for the same $\beta$ and $\mu_0$, this indicates that a sequence
    of two transitions (interpreted as entrance into and exit out of the cone)
    could be identified. For $\beta=3.75$, 3.8 and 3.9 the critical hopping
    parameter for $T=0$ (found by the ETM Collaboration [24] at $\mu_0=0$)
    is given for comparison. It should be between the lower and upper
    $\kappa_t$ for $\mu_0 \ne 0$.

\begin{table}[t]
\begin{center}
\begin{tabular}{rrrrc}
$\beta$ & $\kth$ & $\mu_t$ & $\kcr(T=0)$ & observable \\
\hline
\hline
3.75 & 0.1656(8)   & 0.005 & 0.1660      & $\chi(P)$ \\
3.75 & 0.1677(24)   & 0.005 &             & $\vert\pi\vert^2$\\
3.75 & 0.1657(5)   & 0.007 &             & $\chi(P)$ \\
3.75 & 0.1675(4)   & 0.007 &             & $\chi\left(\ReL\right)$\\
3.75 & 0.16574(24) & 0.008 &             & $\vert\pi\vert^2$ \\
3.75 & 0.1658(4)   & 0.010 &             & $\chi(P)$ \\
\hline
3.775& 0.1645(1) & 0.005 &               & $\chi(P)$\\
\hline
3.8  & 0.16361(4) & 0.005 & 0.164111     & $\vert\pi\vert^2$\\
3.8  & 0.16621(5) & 0.005 &              & $\chi\left(\ReL\right)$ \\
\hline
3.9  & 0.1597(3) & 0.005 & 0.160856     & $\chi\left(\ReL\right)$ \\
\hline
3.95(10) & 0.17 & 0.005 & & $\chi\left(\ReL\right)$ \\
\hline
4.50(5) & 0.0001 & 0 & \\
\hline
4.45(5) & 0.05   & 0 & \\
\hline
4.43(5) & 0.1    & 0 & \\
\hline\hline
\end{tabular}
\caption{Parameter values for identified transition points.
\label{tab:TransitionPoints}}
\end{center}
\end{table}

\clearpage
\bibliographystyle{apsrev}
\bibliography{pdtm}

\providecommand{\href}[2]{#2}\begingroup\raggedright\begin{thebibliography}{10}

\bibitem{Philipsen:2007rj}
O.~Philipsen, 
Eur.  Phys. J. Special Topics {\bf 152}, 29 (2007),
[\href{http://xxx.lanl.gov/abs/0708.1293}{{\tt arXiv:0708.1293}}].

\bibitem{DeTar:2008qi}
C.~E. DeTar, 
Proc. Sci. {\bf LATTICE2008}, 001 (2008), 
[\href{http://xxx.lanl.gov/abs/0811.2429}{{\tt arXiv:0811.2429}}].

\bibitem{Symanzik:1983dc}
K.~Symanzik, 
Nucl. Phys. {\bf B 226}, 187 (1983).

\bibitem{Symanzik:1983gh}
K.~Symanzik, 
Nucl. Phys. {\bf B 226}, 205 (1983).

\bibitem{Luscher:1984xn}
M.~L{\"{u}}scher and P.~Weisz, 
Commun. Math. Phys. {\bf 97}, 59 (1985).

\bibitem{Sheikholeslami:1985ij}
B.~Sheikholeslami and R.~Wohlert, 
Nucl. Phys. {\bf B 259}, 572 (1985).

\bibitem{Ukita:2006pc}
N.~Ukita {\em et.~al.}, 
Proc. Sci. {\bf LAT2006}, 150 (2006),
[\href{http://xxx.lanl.gov/abs/hep-lat/0610038}{{\tt hep-lat/0610038}}].

\bibitem{Bornyakov:2007zu}
V.~G. Bornyakov {\em et.~al.} (DIK Collaboration), 
Proc. Sci. {\bf LATTICE2007}, 171 (2007), 
[\href{http://xxx.lanl.gov/abs/0711.1427}{{\tt arXiv:0711.1427}}].

\bibitem{Aoki:2006br}
Y.~Aoki, Z.~Fodor, S.~D. Katz, and K.~K. Szabo, 
Phys. Lett. {\bf B 643}, 46 (2006),
[\href{http://xxx.lanl.gov/abs/hep-lat/0609068}{{\tt hep-lat/0609068}}].

\bibitem{Fodor:2007sy}
Z.~Fodor, 
Proc. Sci. {\bf LATTICE2007}, 011 (2007),
[\href{http://xxx.lanl.gov/abs/0711.0336}{{\tt arXiv:0711.0336}}].

\bibitem{Karsch:2007dp}
F.~Karsch, 
Proc. Sci. {\bf CPOD07}, 026 (2007),
[\href{http://xxx.lanl.gov/abs/0711.0656}{{\tt arXiv:0711.0656}}].

\bibitem{Karsch:2007dt}
F.~Karsch, 
Proc. Sci. {\bf LATTICE2007}, 015 (2007),
[\href{http://xxx.lanl.gov/abs/0711.0661}{{\tt arXiv:0711.0661}}].

\bibitem{Creutz:2007rk}
M.~Creutz, 
Proc. Sci. {\bf LATTICE2007}, 007 (2007),
[\href{http://xxx.lanl.gov/abs/0708.1295}{{\tt arXiv:0708.1295}}].

\bibitem{Kronfeld:2007ek}
A.~S. Kronfeld, 
Proc. Sci. {\bf LATTICE2007}, 016 (2007),
[\href{http://xxx.lanl.gov/abs/0711.0699}{{\tt arXiv:0711.0699}}].

\bibitem{Frezzotti:2000nk}
R.~Frezzotti, P.~A. Grassi, S.~Sint, and P.~Weisz (Alpha Collaboration),
J. High Energy Phys. {\bf 08(2001)}, 058 (2001), 
[\href{http://xxx.lanl.gov/abs/hep-lat/0101001}{{\tt hep-lat/0101001}}].

\bibitem{Frezzotti:2003ni}
R.~Frezzotti and G.~C. Rossi, 
J. High Energy Phys. {\bf 08(2004)}, 007 (2004), 
[\href{http://xxx.lanl.gov/abs/hep-lat/0306014}{{\tt hep-lat/0306014}}].

\bibitem{Shindler:2007vp}
A.~Shindler, 
Phys. Rep. {\bf 461}, 37 (2008), 
[\href{http://xxx.lanl.gov/abs/0707.4093}{{\tt arXiv:0707.4093}}].

\bibitem{Farchioni:2004us}
F.~Farchioni {\em et.~al.}, 
Eur. Phys. J. {\bf C 39}, 421 (2005),
[\href{http://xxx.lanl.gov/abs/hep-lat/0406039}{{\tt hep-lat/0406039}}].

\bibitem{Farchioni:2004ma}
F.~Farchioni {\em et.~al.}, 
Nucl. Phys. Proc. Suppl. {\bf 140}, 240 (2005), 
[\href{http://xxx.lanl.gov/abs/hep-lat/0409098}{{\tt hep-lat/0409098}}].

\bibitem{Farchioni:2004fs}
F.~Farchioni {\em et.~al.}, 
Eur. Phys. J. {\bf C 42}, 73 (2005),
[\href{http://xxx.lanl.gov/abs/hep-lat/0410031}{{\tt hep-lat/0410031}}].

\bibitem{Farchioni:2005ec}
F.~Farchioni {\em et.~al.}, 
Proc. Sci. {\bf LAT2005}, 072 (2006),
[\href{http://xxx.lanl.gov/abs/hep-lat/0509131}{{\tt hep-lat/0509131}}].

\bibitem{Farchioni:2005bh}
F.~Farchioni {\em et.~al.}, 
Eur. Phys. J. {\bf C 47}, 453 (2006),
[\href{http://xxx.lanl.gov/abs/hep-lat/0512017}{{\tt hep-lat/0512017}}].

\bibitem{Farchioni:2005tu}
F.~Farchioni {\em et.~al.}, 
Phys. Lett. {\bf B 624}, 324 (2005),
[\href{http://xxx.lanl.gov/abs/hep-lat/0506025}{{\tt hep-lat/0506025}}].

\bibitem{Urbach:2007rt}
C.~Urbach (ETM Collaboration), 
Proc. Sci. {\bf LATTICE2007}, 022 (2007), 
[\href{http://xxx.lanl.gov/abs/0710.1517}{{\tt arXiv:0710.1517}}].

\bibitem{Boucaud:2007uk}
Ph.~Boucaud {\em et.~al.} (ETM Collaboration), 
Phys. Lett. {\bf B 650}, 304 (2007),
[\href{http://xxx.lanl.gov/abs/hep-lat/0701012}{{\tt hep-lat/0701012}}].

\bibitem{Boucaud:2008xu}
Ph.~Boucaud {\em et.~al.} (ETM Collaboration),
Comput.  Phys. Commun. {\bf 179}, 695 (2008),
[\href{http://xxx.lanl.gov/abs/0803.0224}{{\tt arXiv:0803.0224}}].

\bibitem{Dimopoulos:2007fn}
P.~Dimopoulos {\em et.~al.}, 
Proc. Sci. {\bf LATTICE2007}, 241 (2007), 
[\href{http://xxx.lanl.gov/abs/0710.0975}{{\tt arXiv:0710.0975}}].

\bibitem{Dimopoulos:2008sy}
P.~Dimopoulos {\em et.~al.} (ETM Collaboration), 
Proc. Sci. {\bf LATTICE2008}, 103 (2008),
\href{http://xxx.lanl.gov/abs/0810.2873}{{\tt arXiv:0810.2873}}.

\bibitem{Kogut:2006gt}
J.~B. Kogut and D.~K. Sinclair, 
Phys. Rev. {\bf D 73}, 074512 (2006),
[\href{http://xxx.lanl.gov/abs/hep-lat/0603021}{{\tt hep-lat/0603021}}].

\bibitem{Bonati:2009yi}
C.~Bonati, G.~Cossu, A.~D'Alessandro, M.~D'Elia, and A.~Di~Giacomo, 
Proc. Sci. {\bf LATTICE2008}, 252 (2008), 
[\href{http://xxx.lanl.gov/abs/0901.4429}{{\tt arXiv:0901.4429}}].

\bibitem{Ilgenfritz:2006tz}
E.-M. Ilgenfritz {\em et.~al.}, 
Proc. Sci. {\bf LAT2006}, 140 (2006),
[\href{http://xxx.lanl.gov/abs/hep-lat/0610112}{{\tt hep-lat/0610112}}].

\bibitem{Ilgenfritz:2007qr}
E.-M. Ilgenfritz {\em et.~al.}, 
Proc. Sci. {\bf LATTICE2007}, 238 (2007),
[\href{http://xxx.lanl.gov/abs/0710.0569}{{\tt arXiv:0710.0569}}].

\bibitem{Ilgenfritz:2008td}
E.-M. Ilgenfritz {\em et.~al.}, 
Proc. Sci. {\bf LATTICE2008}, 206 (2008), 
[\href{http://xxx.lanl.gov/abs/0809.5228}{{\tt arXiv:0809.5228}}].

\bibitem{DeGrand:1988vx}
T.~A. DeGrand, 
Comput. Phys. Commun. {\bf 52}, 161 (1988).

\bibitem{Hasenbusch:2001ne}
M.~Hasenbusch, 
Phys. Lett. {\bf B 519}, 177 (2001),
[\href{http://xxx.lanl.gov/abs/hep-lat/0107019}{{\tt hep-lat/0107019}}].

\bibitem{Hasenbusch:2003vg}
M.~Hasenbusch, 
%% {\it {Full QCD algorithms towards the chiral limit}},  {\em
Nucl. Phys. Proc. Suppl. {\bf 129}, 27 (2004),
[\href{http://xxx.lanl.gov/abs/hep-lat/0310029}{{\tt hep-lat/0310029}}].

\bibitem{Weingarten:1991ra}
D.~H. Weingarten and J.~C. Sexton, 
%% {\it {Hamiltonian evolution for the hybrid Monte Carlo algorithm}},  
Nucl. Phys. Proc. Suppl. {\bf 26}, 613 (1992).

\bibitem{Urbach:2005ji}
C.~Urbach, K.~Jansen, A.~Shindler, and U.~Wenger, 
Comput.  Phys. Commun. {\bf 174}, 87 (2006),
[\href{http://xxx.lanl.gov/abs/hep-lat/0506011}{{\tt hep-lat/0506011}}].

\bibitem{Jansen:2009xp}
K. Jansen and C. Urbach, 
\href{http://xxx.lanl.gov/abs/0905.3331}{{\tt arXiv:0905.3331}}.

\bibitem{Aoki:1983qi}
S.~Aoki, 
Phys. Rev. {\bf D 30}, 2653 (1984).

\bibitem{Aoki:1997fm}
S.~Aoki, 
Nucl. Phys. Proc. Suppl. {\bf 60A}, 206 (1998),
[\href{http://xxx.lanl.gov/abs/hep-lat/9707020}{{\tt hep-lat/9707020}}].

\bibitem{Ilgenfritz:2003gw}
E.-M. Ilgenfritz, W.~Kerler, M.~M{\"{u}}ller-Preussker, A.~Sternbeck, and
  H.~St{\"{u}}ben, 
Phys. Rev. {\bf D 69}, 074511 (2004),
[\href{http://xxx.lanl.gov/abs/hep-lat/0309057}{{\tt hep-lat/0309057}}].

\bibitem{Aoki:1986ua}
S.~Aoki, 
Phys. Lett. {\bf B 190}, 140 (1987).

\bibitem{Aoki:1987us}
S.~Aoki, 
Nucl. Phys. {\bf B 314}, 79 (1989).

\bibitem{Aoki:1995yf}
S.~Aoki, A.~Ukawa, and T.~Umemura, 
Phys. Rev. Lett. {\bf 76}, 873 (1996), 
[\href{http://xxx.lanl.gov/abs/hep-lat/9508008}{{\tt hep-lat/9508008}}].

\bibitem{Bitar:1996kc}
K.~M. Bitar, 
Phys. Rev. {\bf D 56}, 2736 (1997),
[\href{http://xxx.lanl.gov/abs/hep-lat/9602027}{{\tt hep-lat/9602027}}].

\bibitem{Sternbeck:2003gy}
A.~Sternbeck, E.-M. Ilgenfritz, W.~Kerler, M.~M{\"{u}}ller-Preussker, and
  H.~St{\"{u}}ben, 
Nucl. Phys. Proc. Suppl. {\bf 129}, 898 (2004),
[\href{http://xxx.lanl.gov/abs/hep-lat/0309059}{{\tt hep-lat/0309059}}].

\bibitem{Ilgenfritz:2005ba}
E.-M. Ilgenfritz, W.~Kerler, M.~M{\"{u}}ller-Preussker, A.~Sternbeck, and
  H.~St{\"{u}}ben, 
\href{http://xxx.lanl.gov/abs/hep-lat/0511059}{{\tt hep-lat/0511059}}.

\bibitem{AliKhan:2000iz}
A.~Ali~Khan {\em et.~al.} (CP-PACS Collaboration), 
Phys. Rev. {\bf D 63}, 034502 (2001),
[\href{http://xxx.lanl.gov/abs/hep-lat/0008011}{{\tt hep-lat/0008011}}].

\bibitem{AliKhan:2001ek}
A.~Ali~Khan {\em et.~al.} (CP-PACS Collaboration), 
Phys. Rev. {\bf D 64}, 074510 (2001),
[\href{http://xxx.lanl.gov/abs/hep-lat/0103028}{{\tt hep-lat/0103028}}].

\bibitem{Bornyakov:2004ii}
V.~G. Bornyakov {\em et.~al.} (DIK Collaboration), 
Phys. Rev. {\bf D 71}, 114504 (2005),
[\href{http://xxx.lanl.gov/abs/hep-lat/0401014}{{\tt hep-lat/0401014}}].

\bibitem{Bornyakov:2005dt}
V.~G. Bornyakov {\em et.~al.}, 
Proc. Sci. {\bf LAT2005}, 157 (2006),
[\href{http://xxx.lanl.gov/abs/hep-lat/0509122}{{\tt hep-lat/0509122}}].

\bibitem{Sharpe:1998xm}
S.~R. Sharpe and R.~L. Singleton~Jr., 
Phys. Rev. {\bf D 58}i, 074501 (1998),
[\href{http://xxx.lanl.gov/abs/hep-lat/9804028}{{\tt hep-lat/9804028}}].

\bibitem{Munster:2004am}
G.~M{\"{u}}nster, 
J. High Energy Phys. {\bf 09(2004)}, 035 (2004),
[\href{http://xxx.lanl.gov/abs/hep-lat/0407006}{{\tt hep-lat/0407006}}].

\bibitem{Creutz:1996bg}
M.~Creutz, 
\href{http://xxx.lanl.gov/abs/hep-lat/9608024}{{\tt hep-lat/9608024}}.

\bibitem{Creutz:2007fe}
M.~Creutz, 
Phys. Rev. {\bf D 76}, 054501 (2007),
[\href{http://xxx.lanl.gov/abs/0706.1207}{{\tt arXiv:0706.1207}}].

\bibitem{Sharpe:2004ny}
S.~R. Sharpe and J.~M.~S. Wu, 
Phys. Rev. {\bf D 71}, 074501 (2005),
[\href{http://xxx.lanl.gov/abs/hep-lat/0411021}{{\tt hep-lat/0411021}}].

\bibitem{Sharpe:2006pu}
S.~R. Sharpe, 
\href{http://xxx.lanl.gov/abs/hep-lat/0607016}{{\tt hep-lat/0607016}}.

\bibitem{Blum:1994eh}
T.~Blum {\em et.~al.}, 
Phys. Rev. {\bf D 50}, 3377 (1994),
[\href{http://xxx.lanl.gov/abs/hep-lat/9404006}{{\tt hep-lat/9404006}}].

\bibitem{Jansen:2003ir}
K.~Jansen, A.~Shindler, C.~Urbach, and I.~Wetzorke (XLF Collaboration), 
Phys. Lett. {\bf B 586}, 432 (2004), 
[\href{http://xxx.lanl.gov/abs/hep-lat/0312013}{{\tt hep-lat/0312013}}].

\bibitem{ETMC:private}
Carsten Urbach (ETM Collaboration) (private communication).

\bibitem{Cheng:2006qk}
M.~Cheng {\em et.~al.}, 
Phys.  Rev. {\bf D 74}, 054507 (2006),
[\href{http://xxx.lanl.gov/abs/hep-lat/0608013}{{\tt hep-lat/0608013}}].

\bibitem{Philipsen:2008gq}
O.~Philipsen and L.~Zeidlewicz, 
\href{http://xxx.lanl.gov/abs/0812.1177}{{\tt arXiv:0812.1177}}.

\bibitem{Bornyakov:2009qh}
V.G.~Bornyakov {\em et al.}, 
\href{http://xxx.lanl.gov/abs/0910.2392}{{\tt arXiv:0910.2392}}.

\bibitem{Baron:2008xa}
R.~Baron {\em et al.} (ETM Collaboration), 
Proc. Sci. {\bf LATTICE2008}, 094 (2008),
[\href{http://xxx.lanl.gov/abs/0810.3807}{{\tt arXiv:0810.3807}}].

\end{thebibliography}\endgroup
\end{document}